\def\figa{\begin{figure*}[!t]
    \begin{center}
      \epsscale{1.1}
      \plotone{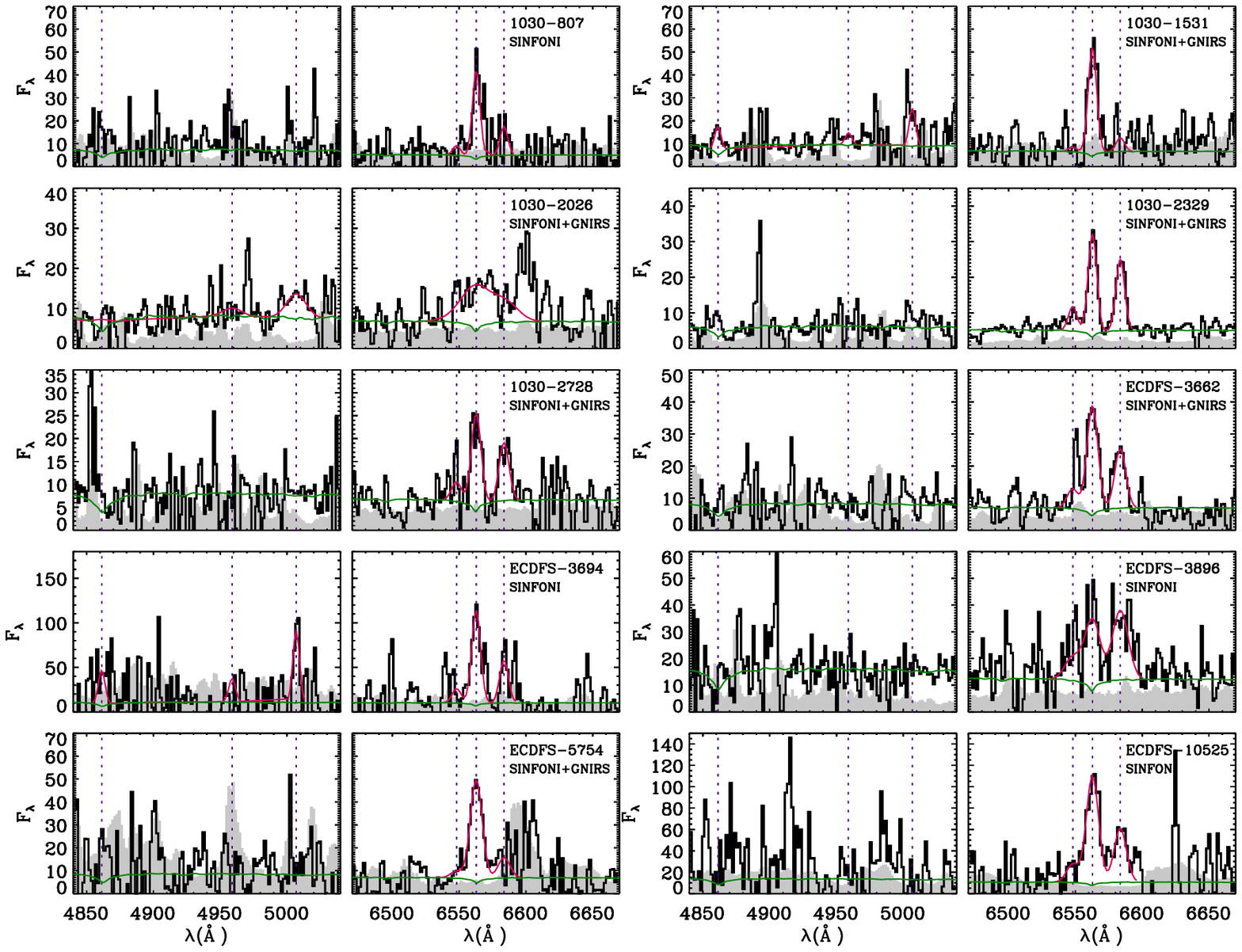}
    \end{center}
    \figcaption{One dimensional spectra in the wavelength region
      around \hb\ and \ot\ $\lambda\lambda$4959, 5007, and \ha\ and
      \nt\ $\lambda \lambda$6548, 6583 of ten $K$-selected emission
      line galaxies at $2.0<z<2.7$. The wavelength is presented in
      rest-frame and the flux is given in $10^{-19}\,\rm ergs\,s^{-1}
      cm^{-2} \AA^{-1}$. The vertical dotted lines present the
      positions of the expected \hb, \ot\ $\lambda \lambda$ 4959,
      5007, \nt\ $\lambda$6548, \ha\ and \nt\ $\lambda$6583 lines. The
      red line presents the best fit to the three emission lines. The
      green line is the best continuum fit to the low-resolution GNIRS
      spectra. Gray shaded areas present the noise spectrum. For
      several galaxies we combined the SINFONI with the GNIRS spectra
      for reasons explained in the text.\label{spectra}}
  \end{figure*}
}

\def\figb{\begin{figure*}[!t]
    \begin{center}
      \epsscale{1.1}
      \plotone{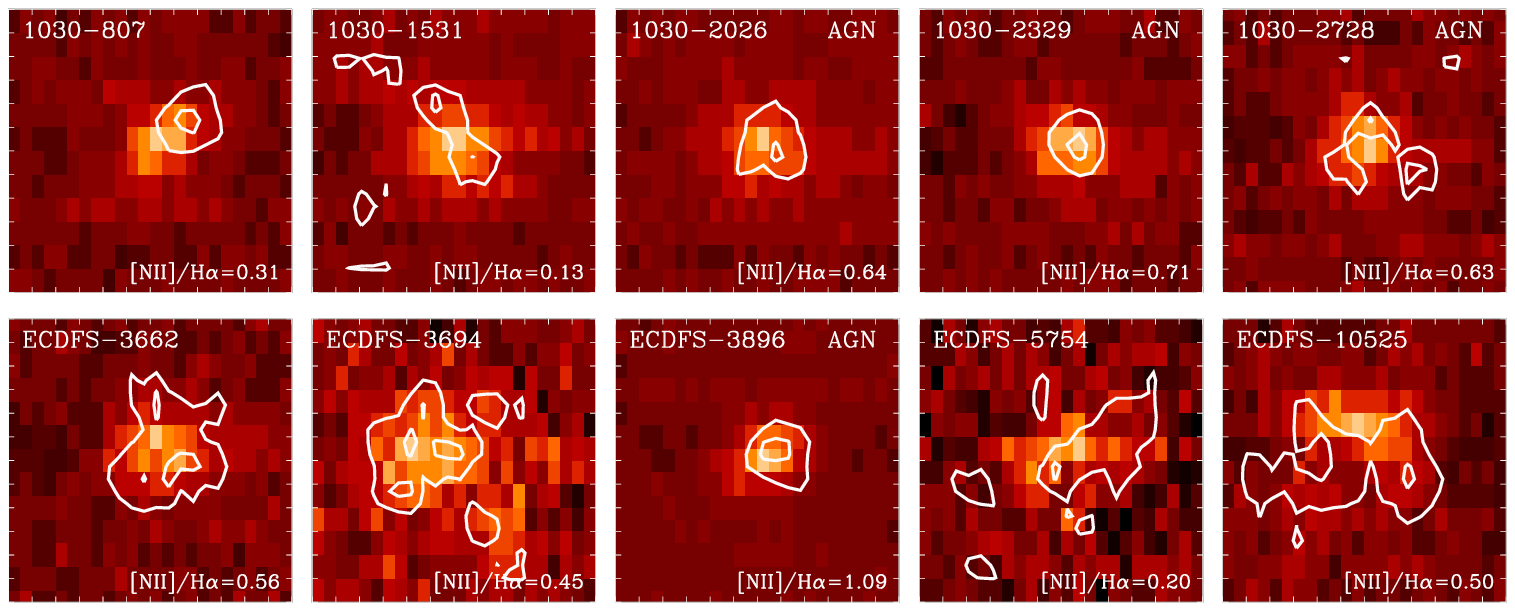}
    \end{center}
    \caption{3\arcsec\ by 3\arcsec\ images ($\sim$25 by 25 kpc) of the
      continuum ({\it color scale}) and line ({\it contours})
      emission. The continuum emission is derived from the median
      collapsed H+K SINFONI cubes, excluding wavelength regions with
      low atmospheric transmission or strong sky emission. The
      linemaps include both the \ha\ and \nt\ $\lambda$6583
      emission. The contours represent the 50\% and 90\% of the
      maximum line emission in the galaxy. Three of the four AGN
      candidates exhibit compact line and continuum
      emission.\label{linemap}}
  \end{figure*}
}

\def\figc{\begin{figure*}[!t]
    \begin{center}
      \epsscale{1.0} 
      \plotone{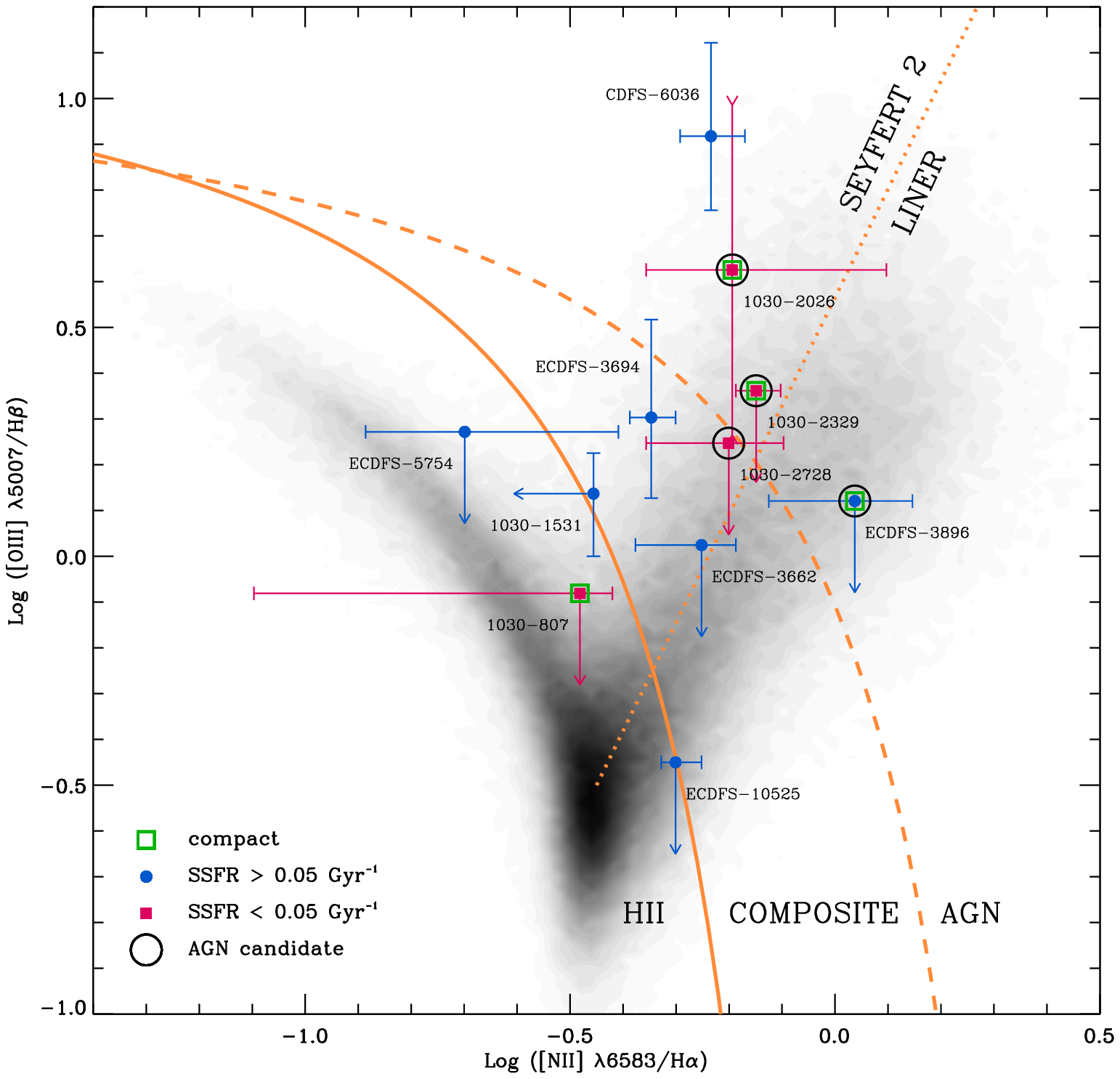} 
      \figcaption{Diagnostic diagram for spectral classification of
        AGNs and star-forming galaxies. The gray-scale presents the
        locus of $\sim$400 000 SDSS galaxies \citep{ka03b,tr04}. The
        orange solid line is the empirical division between galaxies
        for which the line emission originates from H\,{\sc ii}
        regions and AGNs for the SDSS galaxies by \cite{ka03b}. The
        orange dashed line presents the theoretical upper limit by
        \cite{ke01} for star forming galaxies. Galaxies between these
        two dividing curves are classified as composite H\,{\sc
          ii}-AGN galaxies by \cite{ke06}. The orange dotted line
        presents the division between Seyfert 2s and LINERs by
        \cite{ka03b}. The red filled squares present galaxies with a
        specific SFR (derived from modeling the continuum spectra)
        less than 0.05 Gyr$^{-1}$, and the blue filled circles
        galaxies with higher specific SFRs ($>0.05\,\rm
        Gyr^{-1}$). Furthermore, the green squares indicate galaxies
        with compact line emission. The galaxies with black circles
        are identified as AGN candidates, based on their \nt/\ha\
        ratios, spatial extent of the line emission, and ancillary
        data. Further details are in the text. All upper limits are
        2$\sigma$ and the error-bars are all 1$\sigma$. For 1030-2026
        we have both a 2$\sigma$ upper and lower limit.
        \label{ratios}}
    \end{center}
  \end{figure*}
}

\def\figd{
  \begin{figure}[!t]
    \begin{center}
      \epsscale{1.1} 
      \plotone{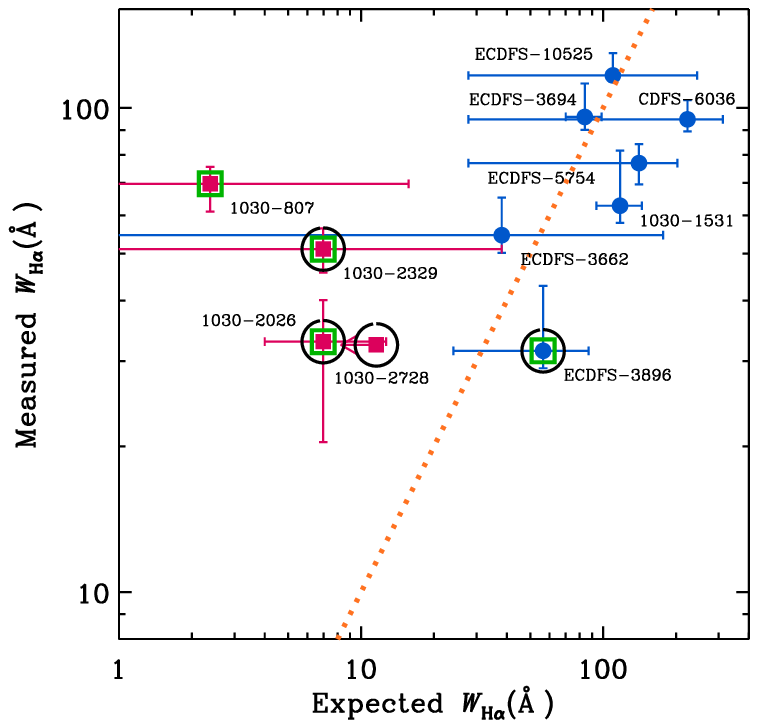} 
      \figcaption{The measured \wha\ vs. the expected \wha\ derived
        from the spectral continuum shape (see Figure~\ref{SEDS}) and
        the \cite{ke98} relation between SFR and \ha\ luminosity. In
        case H\,{\sc ii} regions are the only contributors to the line
        emission, a galaxy falls on the expected 1-to-1 relation ({\it
          dotted line}). Additional extinction towards star forming
        regions moves a galaxy downwards of the relation. Objects that
        fall above the dotted line most likely have another ionization
        source contributing to the line emission as well. The symbols
        are similar as in Figure~\ref{ratios}.\label{wha}}
    \end{center}
  \end{figure}
}

\def\fige{
  \begin{figure}[!t]
    \begin{center}
      \includegraphics[scale=0.85]{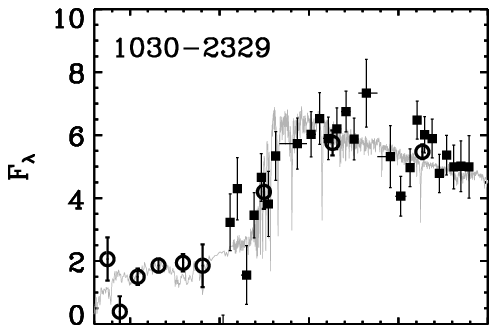}\hspace{-0.1in}
      \includegraphics[scale=0.85]{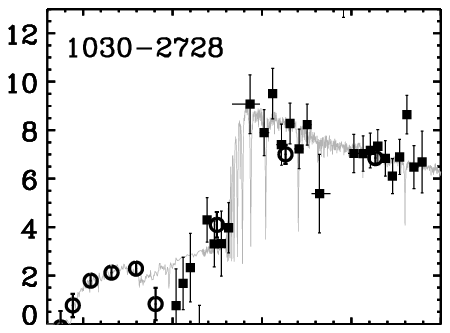}\\\vspace{-0.1in}
      \includegraphics[scale=0.85]{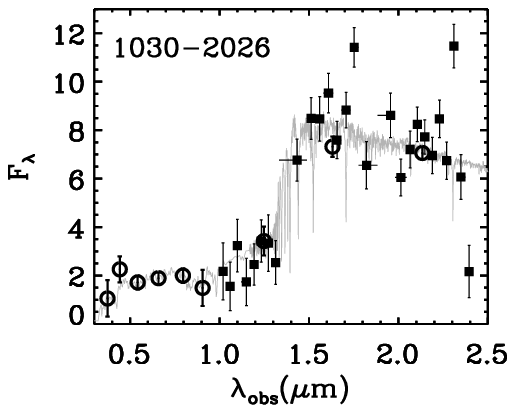}\hspace{-0.1in}
      \includegraphics[scale=0.85]{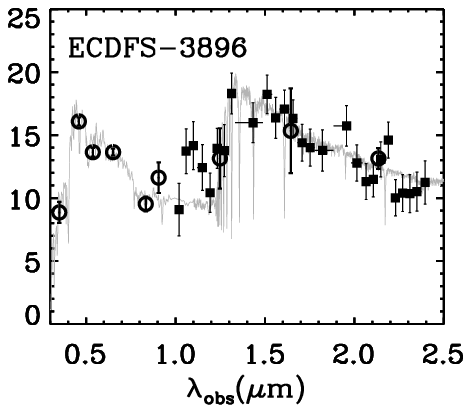}\\
    \end{center}
    \figcaption{Binned ``low resolution'' spectra ({\it filled
        squares}) and optical-to-NIR photometry ({\it open circles})
      of the 4 AGN host galaxies. The best-fit stellar population
      models are drawn in gray. The detection of strong Balmer and/or
      4000 \AA\ breaks in three of the four galaxies implies that the
      continuum emission in these galaxies is dominated by stellar
      light. The SED of ECDFS-3896 indicates active star formation in
      this galaxy, while the other three galaxies are best fitted by
      evolved stellar populations. \label{SEDS}}
  \end{figure}
}

\def\figf{
  \begin{figure*}[!t]
    \begin{center}
      \epsscale{0.8} \plotone{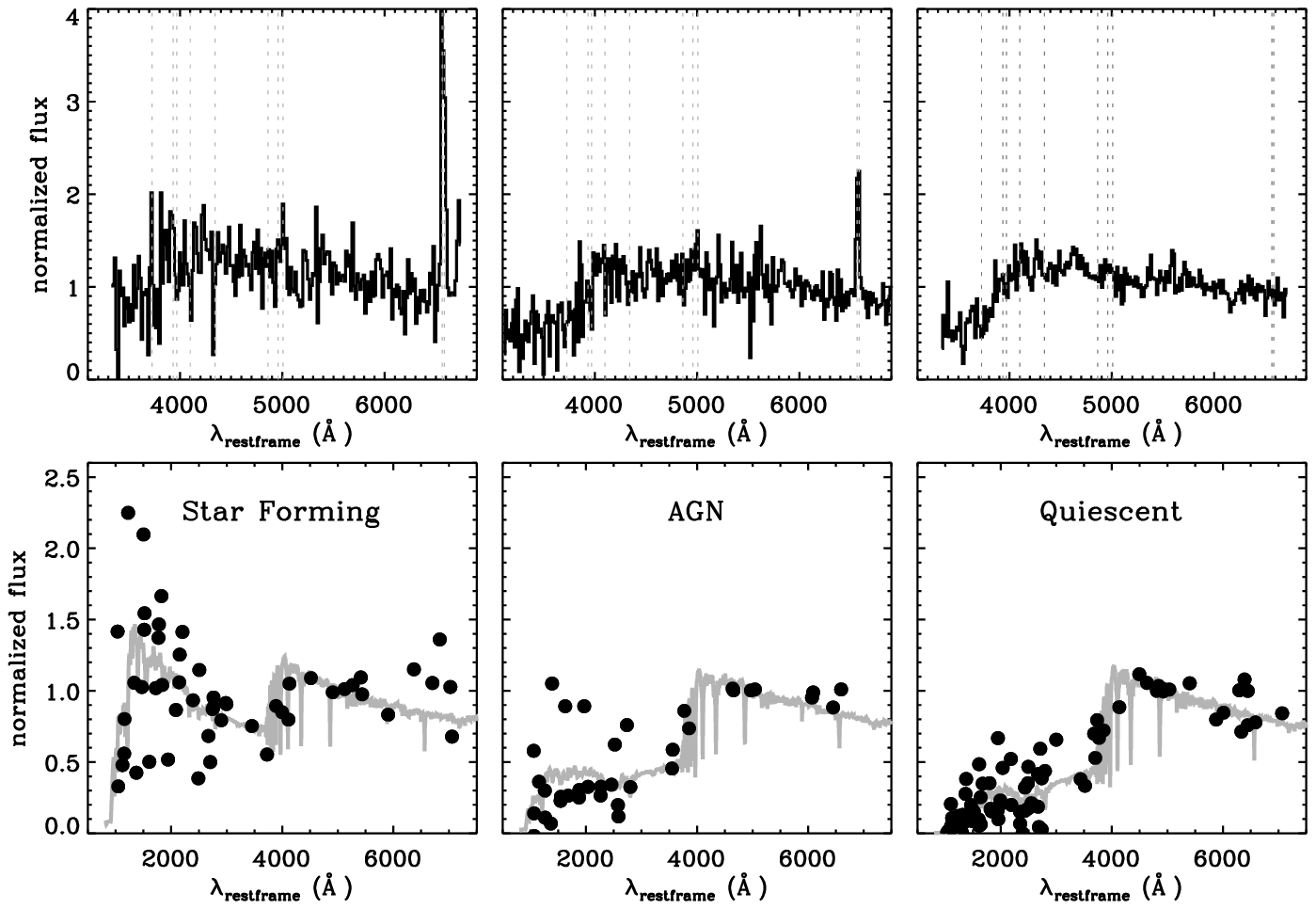} \figcaption{Stacked GNIRS
        spectra and composite broadband SEDs of the three different
        classes of galaxies in our sample: the nine quiescent galaxies
        without detected emission lines, the four AGN candidates, and
        the remaining seven emission line galaxies. The individual
        GNIRS spectra were normalized in the wavelength interval
        between 4000 \AA\ and 6000 \AA, and the individual broadband
        SEDs were set to unity at a wavelength of 5000 \AA. The dotted
        vertical lines present from left to right the location of
        [O\,{\sc ii}] $\lambda$3727, Ca(H), Ca(K), H$\delta$,
        H$\gamma$, H$\beta$, \ot\ $\lambda$4959, \ot\ $\lambda$5007,
        \ha\, \nt\ $\lambda$6583. The gray lines in the lower panels
        show the average of the individual best-fit stellar population
        models to the spectra and broadband photometry for each
        class. \label{stacks}}
    \end{center}
  \end{figure*}
}

\def\figg{
  \begin{figure*}[!t]
    \begin{center}
      \epsscale{1.}  \plottwo{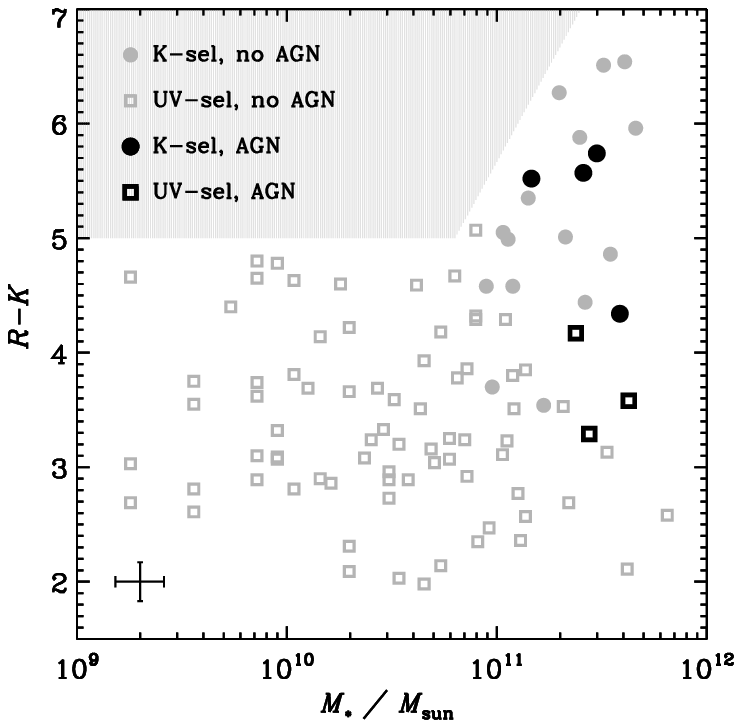}{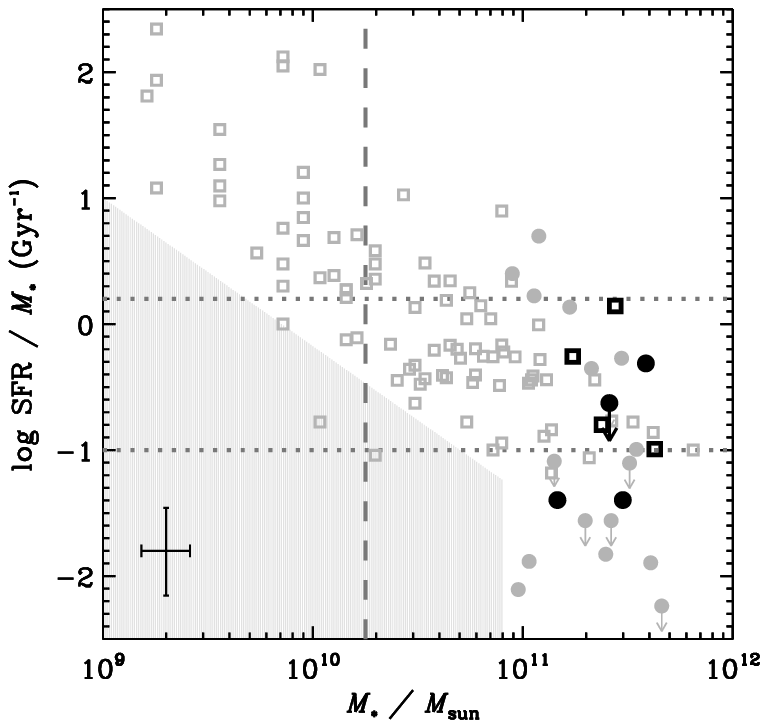} \figcaption{$R-K$ and
        the specific SFR (as derived from the stellar continua)
        vs. the stellar mass for both the UV-selected galaxies by
        \cite{er06a,er06b,er06c} and our spectroscopic $K$-selected
        sample at $2.0<z<2.7$. The masses for the UV and $K$-selected
        samples are derived from modeling the broadband photometry and
        the stellar continua respectively, assuming a \cite{sa55}
        IMF. The galaxies in which AGNs are identified are indicated
        in black. The shaded gray areas in both diagrams are empty
        probably because of incompleteness effects. We can roughly
        remove the bias towards AGN with more massive black holes, by
        normalizing the star-formation rate by the stellar mass,
        assuming that it scales with the black hole mass. Considering
        a line of constant specific SFR the line emission originating
        from accretion onto the black hole should be equally easy to
        detect as the line emission from star formation. Thus, AGNs
        with the same normalized accretion rate should have been
        detected in the area between the dotted lines and to the right
        of the shaded region. This figure illustrates that black hole
        accretion is more effective at the high-mass end at
        $z\sim2.3$. \label{masses}}
    \end{center}
  \end{figure*}
}

\def\figh{\begin{figure}[!t]
    \begin{center}
      \epsscale{1.1} 
      \plotone{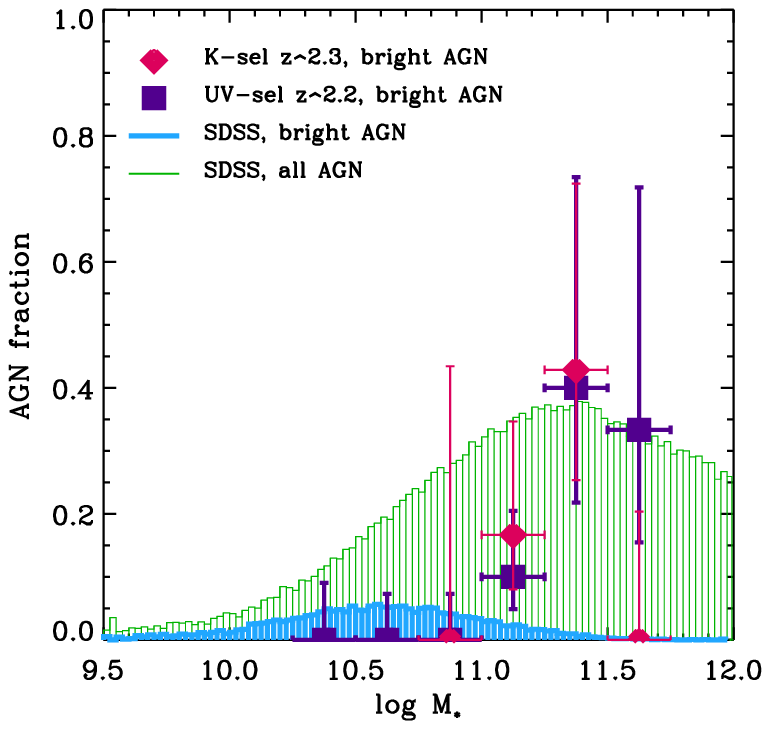} 
      \figcaption{Fraction of actively
        accreting AGN among galaxies in different mass bins at
        different redshifts. The red diamonds present the AGN fraction
        for the 20 $K$-selected galaxies at $2.0<z<2.7$. The purple
        squares present the AGN fraction among the UV-selected
        galaxies at $z\sim2.2$ by \cite{er06a}. The rapid fall off of
        both histograms may be due to selection effects. The blue
        histogram is the fraction of AGNs in the SDSS sample
        \citep{ka03a} with similar normalized accretion rates as the
        AGNs in our $K$-selected sample. The green histogram are all
        AGNs in the SDSS, as defined in the text.  All stellar masses
        are derived from stellar population modeling, and converted to
        a \cite{sa55} IMF. This diagram illustrates that actively
        accreting AGNs reside in more massive galaxies at higher
        redshift.\label{fractions}}
    \end{center}
  \end{figure}
}

\def\figi{
  \begin{figure*}[!t]
    \begin{center}
      \epsscale{1.} 
      \plottwo{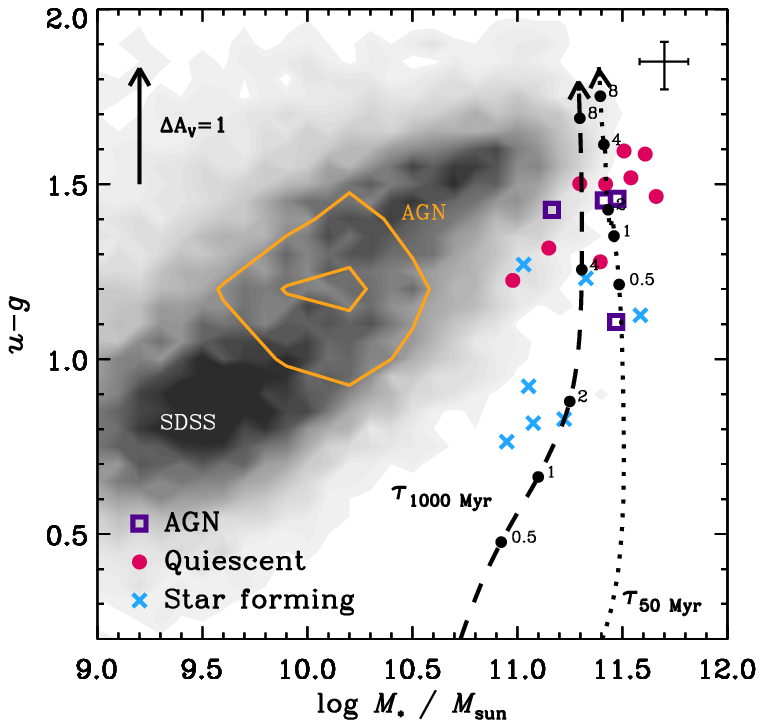}{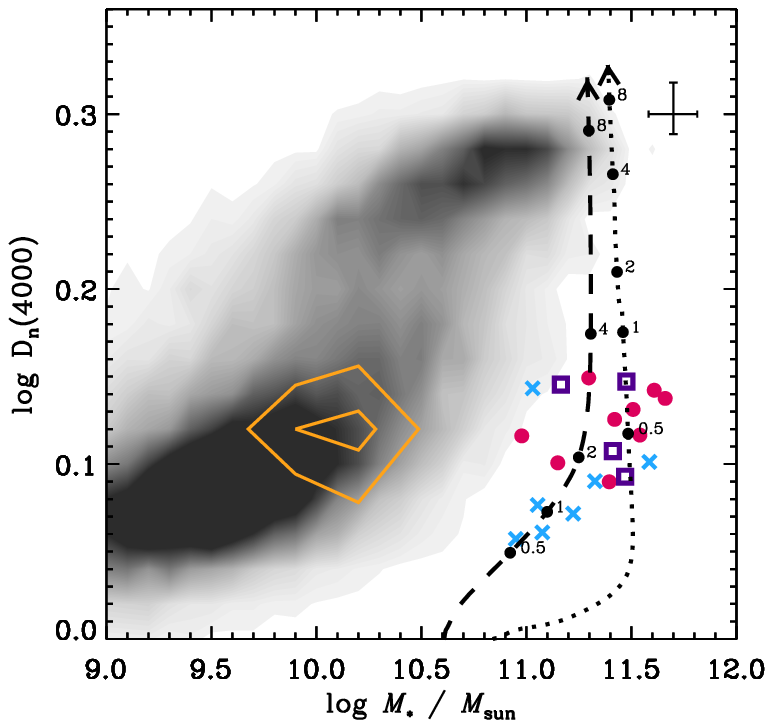} 
      \figcaption{Rest-frame
        $u-g$ color and $D_{\rm n}(4000)$ vs. stellar mass for the
        $K$-selected $2.0<z<2.7$ galaxy sample. Both properties ($u-g$
        and $D_{\rm n}(4000)$) are derived from the best fits to the
        rest-frame optical continuum spectra in combination with the
        rest-frame UV broad-band photometry. The errors are derived
        from the 68\% best-fits of 200 Monte Carlo simulations. The
        gray scale presents the low-redshift catalogue of galaxies
        extracted from the SDSS by \cite{bl05}. The photometric masses
        \citep{ka03a} are corrected for the difference in assumed IMF
        with the $K$-selected sample. The actively accreting AGNs
        within the SDSS sample, as defined in \S~\ref{down} are
        represented by the orange contours. The evolution of a stellar
        population for two different SFHs with different decaying
        times ($\tau$) are drawn, assuming a formation redshift of
        2.8. The ages are indicated in Gyr. These diagrams may suggest
        that both at low and high redshift the AGN activity peaks
        during the post-starburst phase, when a galaxy is transforming
        from a star-forming into a quiescent galaxy.\label{colmag}}
    \end{center}
  \end{figure*}
}

\def\taba{\begin{deluxetable}{llllrrc}
    \tabletypesize{\scriptsize} 
    \tablecaption{Sample\label{obs}}
    \tablewidth{0pt}
    \tablehead{ \colhead{id} & \colhead{$z_{\rm H\alpha}$} & 
      \colhead{$K_s$} & \colhead{$R$} & 
      \colhead{t$_{\rm S}$\tablenotemark{a}} & 
      \colhead{t$_{\rm G}$\tablenotemark{b}} & \colhead{$L_{\rm 0.5-8.0 keV}$} \\
      \colhead{} & \colhead{} & \colhead{} & \colhead{} & \colhead{min} &
      \colhead{min} & \colhead{ergs\,s$^{-1}$}}
    \startdata
    1030-807 & 2.367 & 19.72 & 24.77 & 80 & 120 & $<3 \times 10^{43}$\\
    1030-1531 & 2.613 & 19.38 & 22.92 & 110 & 80 & $<3 \times 10^{43}$ \\
    1030-2026 & 2.512 & 19.48 & 25.22 & 120 & 120 & $<6 \times 10^{43}$ \\
    1030-2329 & 2.236 & 19.72 & 25.24 & 80 & 120 & $<1 \times 10^{43}$ \\
    1030-2728 & 2.504 & 19.52 & 25.09 & 110 & 120 & $<2 \times 10^{43}$ \\
    ECDFS-3662 & 2.350 & 19.20 & 24.29 & 60 & 100 & $<1 \times 10^{42}$ \\
    ECDFS-3694 & 2.122 & 18.90 & 23.60 & 70 & 190 & $<1 \times 10^{42}$ \\
    ECDFS-3896 & 2.308 & 18.82 & 23.02 & 60 & 60 & $<5 \times 10^{42}$ \\
    ECDFS-5754 & 2.037 & 19.36 & 23.54 & 70 & 150 & $<5 \times 10^{42}$ \\
    ECDFS-10525 & 2.024 & 19.15 & 22.70 & 90 & 90 & $<5 \times 10^{42}$ \\
    CDFS-6036\tablenotemark{c} & 2.225 & 19.12 & 22.87 & - & 92 & $2.7 \times 10^{42}$ \\ 
    \enddata
    \tablenotetext{a}{Integration times for SINFONI.}
    \tablenotetext{b}{Integration times for GNIRS.}
    \tablenotetext{c}{Observations presented by \cite{vd05} and
      \cite{kr06a}. The $R$ magnitude is adopted from \cite{da04a}.}
  \end{deluxetable}
}

\def\tabb{\begin{deluxetable*}{l r r r r l l}
    \tabletypesize{\scriptsize} \tablecaption{Stellar population
      properties of the emission line galaxies\label{cont}}
    \tablewidth{0pt} \tablehead{ \colhead{id} & \colhead{$\tau$} &
      \colhead{age} & \colhead{$A_V$} & \colhead{$M_{*}$} &
      \colhead{SFR} &  \colhead{SFR/$M_{*}$} \\
      \colhead{} & \colhead{Gyr} & \colhead{Gyr} & \colhead{mag} &
      \colhead{$10^{11} M_{\odot}$} & \colhead{$M_{\odot}$\,yr$^{-1}$} &
      \colhead{$10^{-2} \rm \, Gyr^{-1}$} 
    } 
    \startdata 
    1030-807&
    0.12$_{- 0.11}^{+ 0.03}$ &
    0.81$_{- 0.60}^{+ 0.10}$ &
    0.0$_{-  0.0}^{+  1.6}$ &
    1.1$_{-  0.1}^{+  0.7}$ &
    1$_{-    1}^{+   32}$ &
    1$_{-    1}^{+   18}$ \\
    1030-1531&
    0.65$_{- 0.40}^{+ 9.35}$ &
    0.57$_{- 0.17}^{+ 0.57}$ &
    0.8$_{-  0.1}^{+  0.1}$ &
    1.7$_{-  0.3}^{+  0.6}$ &
    227$_{-   59}^{+   72}$ &
    136$_{-   37}^{+   51}$ \\
    1030-2026&
    0.15$_{- 0.03}^{+ 0.05}$ &
    0.81$_{- 0.24}^{+ 0.33}$ &
    0.8$_{-  0.5}^{+  0.5}$ &
    3.0$_{-  0.7}^{+  0.7}$ &
    12$_{-    7}^{+   20}$ &
    4$_{-    1}^{+    5}$ \\
    1030-2329&
    0.15$_{- 0.14}^{+ 0.05}$ &
    0.81$_{- 0.60}^{+ 0.33}$ &
    0.7$_{-  0.4}^{+  1.4}$ &
    1.5$_{-  0.3}^{+  0.8}$ &
    5$_{-    5}^{+   78}$ &
    4$_{-    4}^{+   40}$ \\
    1030-2728&
    0.02$_{- 0.01}^{+ 0.08}$ &
    0.29$_{- 0.08}^{+ 0.28}$ &
    1.3$_{-  0.4}^{+  0.5}$ &
    2.6$_{-  0.4}^{+  0.8}$ &
    0$_{-    0}^{+   34}$ &
    0$_{-    0}^{+   10}$ \\
    ECDFS-3662&
    0.08$_{- 0.07}^{+ 9.92}$ &
    0.29$_{- 0.18}^{+ 0.52}$ &
    1.5$_{-  0.9}^{+  0.6}$ &
    2.1$_{-  0.5}^{+  1.3}$ &
    94$_{-   93}^{+  627}$ &
    44$_{-   43}^{+  244}$ \\
    ECDFS-3694&
    10.00$_{- 7.50}^{+ 0.00}$ &
    2.40$_{- 0.79}^{+ 0.60}$ &
    1.3$_{-  0.1}^{+  0.1}$ &
    3.9$_{-  0.7}^{+  0.7}$ &
    187$_{-   40}^{+   45}$ &
    48$_{-   16}^{+   25}$ \\
    ECDFS-3896&
    0.20$_{- 0.10}^{+ 0.10}$ &
    0.51$_{- 0.10}^{+ 0.06}$ &
    1.0$_{-  0.4}^{+  0.2}$ &
    2.9$_{-  1.0}^{+  0.1}$ &
    157$_{-  113}^{+  122}$ &
    53$_{-   31}^{+   41}$ \\
    ECDFS-5754&
    10.00$_{- 9.99}^{+ 0.00}$ &
    0.72$_{- 0.67}^{+ 0.72}$ &
    1.3$_{-  0.2}^{+  0.2}$ &
    1.1$_{-  0.6}^{+  0.4}$ &
    188$_{-  150}^{+  112}$ &
    167$_{-   89}^{+  223}$ \\
    ECDFS-10525&
    0.04$_{- 0.03}^{+ 9.96}$ &
    0.10$_{- 0.05}^{+ 0.10}$ &
    1.3$_{-  0.2}^{+  0.2}$ &
    0.9$_{-  0.2}^{+  0.2}$ &
    223$_{-  166}^{+  364}$ &
    251$_{-  173}^{+  325}$ \\
    CDFS-6036&
    0.65$_{- 0.64}^{+ 9.35}$ &
    0.20$_{- 0.15}^{+ 0.37}$ &
    1.7$_{-  0.3}^{+  0.2}$ &
    1.2$_{-  0.4}^{+  0.4}$ &
    594$_{-  532}^{+  384}$ &
    498$_{-  420}^{+  544}$ \\
    
    \enddata
    \tablecomments{The stellar population properties are derived from
      fitting the low-resolution continuum spectra and optical
      photometry by stellar population models. The errors present the 68\%
      confidence intervals derived using 200 Monte Carlo simulations.}
  \end{deluxetable*}
}

\def\tabc{\begin{deluxetable*}{l l l l l l l l l}
    \tabletypesize{\scriptsize} 
    \tablecaption{Emission line modeling results\label{model}}
    \tablewidth{0pt}
    \tablehead{ \colhead{id} & \colhead{$\sigma$\tablenotemark{a}} & 
      \colhead{\wha\tablenotemark{b}} & \colhead{$W_{\rm [NII]}$} & 
      \colhead{$W_{\rm H\beta}$} & \colhead{$W_{\rm [OIII]}$} & 
      \colhead{\nt/\ha} & \colhead{\ot/\hb\tablenotemark{c}} & 
      \colhead{\ot/\hb\tablenotemark{d}}}
    \startdata
    1030-807&  115$^{+   21}_{-   52}$ &
    61.7$^{+ 13.1}_{-  8.6}$ &
    18.7$^{+  3.1}_{- 14.5}$ &
    13.0$^{+  6.4}_{-  5.9}$ &
    -1.1$^{+  2.6}_{-  2.0}$ &
    0.33$^{+ 0.05}_{- 0.25}$ &
    $<$ 0.83   &
    $<$ 0.86  \\
    1030-1531&   72$^{+   54}_{-   47}$ &
    62.8$^{+ 18.8}_{-  4.9}$ &
    7.6$^{+  6.1}_{- 12.2}$ &
    9.5$^{+  1.6}_{-  1.2}$ &
    11.0$^{+  2.7}_{-  3.1}$ &
    $<$ 0.35   &
    1.37$^{+ 0.31}_{- 0.37}$ &
    $<$ 1.99  \\
    1030-2026&  434$^{+  117}_{-   64}$ &
    32.9$^{+  7.2}_{- 12.5}$ &
    21.0$^{+ 11.1}_{-  6.4}$ &
    2.0$^{+  3.2}_{-  5.2}$ &
    14.3$^{+  4.6}_{-  1.1}$ &
    0.64$^{+ 0.61}_{- 0.20}$ &
    $>$ 1.84   &
    $<$ 9.70  \\
    1030-2329&   80$^{+   19}_{-   13}$ &
    51.1$^{+  5.3}_{-  5.4}$ &
    34.5$^{+  4.7}_{-  4.1}$ &
    7.6$^{+  2.8}_{-  1.4}$ &
    3.2$^{+  5.2}_{-  2.5}$ &
    0.71$^{+ 0.08}_{- 0.06}$ &
    $<$ 2.30   &
    $<$ 2.51  \\
    1030-2728&  114$^{+   29}_{-   39}$ &
    32.4$^{+  6.9}_{-  4.9}$ &
    19.4$^{+  5.1}_{-  5.3}$ &
    3.7$^{+  2.1}_{-  1.9}$ &
    0.9$^{+  2.5}_{-  1.8}$ &
    0.63$^{+ 0.17}_{- 0.19}$ &
    - &
    $<$ 1.77  \\
    ECDFS-3662&  141$^{+   32}_{-   20}$ &
    54.6$^{+ 10.7}_{-  4.4}$ &
    29.4$^{+  6.8}_{-  6.8}$ &
    2.4$^{+  5.3}_{-  1.5}$ &
    1.8$^{+  0.4}_{-  3.9}$ &
    0.56$^{+ 0.09}_{- 0.14}$ &
    - &
    $<$ 1.06  \\
    ECDFS-3694&  139$^{+   19}_{-   19}$ &
    95.8$^{+ 16.5}_{-  5.7}$ &
    41.5$^{+  9.0}_{-  4.5}$ &
    28.7$^{+ 17.1}_{-  9.1}$ &
    48.9$^{+ 10.7}_{-  3.4}$ &
    0.45$^{+ 0.05}_{- 0.04}$ &
    2.01$^{+ 1.28}_{- 0.67}$ &
    $<$ 4.88  \\
    ECDFS-3896&  265$^{+   10}_{-   64}$ &
    31.5$^{+ 11.4}_{-  2.5}$ &
    34.1$^{+ 11.4}_{-  6.6}$ &
    5.7$^{+  1.9}_{-  5.0}$ &
    2.2$^{+  1.9}_{-  2.3}$ &
    1.09$^{+ 0.31}_{- 0.34}$ &
    - &
    $<$ 1.32  \\
    ECDFS-5754&  160$^{+   17}_{-   24}$ &
    76.9$^{+  7.3}_{-  7.4}$ &
    15.4$^{+ 14.3}_{-  5.7}$ &
    22.3$^{+  9.8}_{- 11.7}$ &
    5.1$^{+  7.7}_{-  6.0}$ &
    0.20$^{+ 0.19}_{- 0.07}$ &
    $<$ 1.87   &
    $<$ 2.21  \\
    ECDFS-10525&  202$^{+   29}_{-    0}$ &
    116.7$^{+ 13.1}_{-  3.5}$ &
    57.9$^{+  9.7}_{-  1.4}$ &
    7.8$^{+ 12.7}_{-  8.0}$ &
    -5.2$^{+  5.4}_{-  3.5}$ &
    0.50$^{+ 0.06}_{- 0.03}$ &
    - &
    $<$ 0.35  \\
    CDFS-6036\tablenotemark{e}& - &
    99$^{+ 10}_{- 10}$ &
    60$^{+  6}_{-  6}$ &
    12$^{+ 5}_{-  5}$ &
    99$^{+  10}_{-  10}$ &
    0.58$^{+ 0.09}_{- 0.08}$ &
    8.28$^{+ 5.39}_{- 2.53}$ &
    - \\
    \enddata
    \tablenotetext{a}{In units of km\,s$^{-1}$, corrected for
      instrumental resolution. For ECDFS-3694 and ECDFS-10525 the 
      velocity gradient is removed from the velocity dispersion, 
      to deblend the emission lines.}
    \tablenotetext{b}{All equivalent widths are in rest-frame, corrected
      for Balmer absorption, and given in \AA.}
    \tablenotetext{c}{Derived from the emission line modeling.}
    \tablenotetext{d}{Derived from the lower limit on \hb\ and the upper
      limit on \ot. The lower limit on \hb\ is estimated using the lower
      limit on \ha, the intrinsic ratio between \ha\ and \hb\, and the
      modeled upper limit on the continuum attenuation. Furthermore we assume
      extra attenuation towards H\,{\sc ii} regions.}
    \tablenotetext{e}{Measured by \cite{vd05}}
    \tablecomments{The errors present the best 68\% confidence intervals 
      derived using 500 Monte Carlo simulations. All upper and lower limits 
      are 2$\sigma$. The emission lines ratios present the flux ratios. }
  \end{deluxetable*}
}

\documentclass[aaspp4,11pt,preprint,apjfonts]{emulateapj}

\slugcomment{Accepted for publication in the Astrophysical Journal}
\shorttitle{AGN in Massive $\lowercase{z}\sim2.3$ Galaxies}
\shortauthors{Kriek et al.}

\newcommand{\ha}{H$\alpha$}
\newcommand{\hb}{H$\beta$}
\newcommand{\av}{$A_V$}
\newcommand{\wha}{$W_{\rm H\alpha}$}
\newcommand{\nt}{[N\,{\sc ii}]}
\newcommand{\ot}{[O\,{\sc iii}]}

\begin{document}
  
\title{The Origin of Line Emission in Massive $\lowercase{z}\sim2.3$
  Galaxies: Evidence for Cosmic Downsizing of AGN Host
  Galaxies\altaffilmark{1,2}}

\author{Mariska Kriek\altaffilmark{3,4},
  Pieter G. van Dokkum\altaffilmark{4,5}, 
  Marijn Franx\altaffilmark{3},  
  Garth D. Illingworth\altaffilmark{6},
  Paolo Coppi\altaffilmark{4,5},
  Natascha M. F{\"o}rster Schreiber\altaffilmark{7},
  Eric Gawiser\altaffilmark{4,5,8},
  Ivo Labb\'e\altaffilmark{9},
  Paulina Lira\altaffilmark{10},
  Danilo Marchesini\altaffilmark{4}, 
  Ryan Quadri\altaffilmark{4}, 
  Gregory Rudnick\altaffilmark{11},
  Edward N. Taylor\altaffilmark{2},
  C. Megan Urry\altaffilmark{5},
  Paul P. van der Werf\altaffilmark{3}}

\email{mariska@strw.leidenuniv.nl}

\altaffiltext{1}{Based on observations collected at the European
  Southern Observatory, Paranal, Chile (076.A-0464 and 076.A-0718)}

\altaffiltext{2}{Based on observations obtained at the Gemini
  Observatory, which is operated by the Association of Universities for
  Research in Astronomy, Inc., under a cooperative agreement with the
  NSF on behalf of the Gemini partnership.}

\altaffiltext{3}{Leiden Observatory, Leiden University, PO Box 9513,
  2300 RA Leiden, The Netherlands}

\altaffiltext{4}{Department of Astronomy, Yale University, New Haven, 
  CT 06520}

\altaffiltext{5}{Yale Center for Astronomy and Astrophysics, Yale
  University, New Haven, CT 06520}

\altaffiltext{6}{UCO/Lick Observatory, University of California, Santa
  Cruz, CA 95064}

\altaffiltext{7}{Max-Planck-Institut f\"ur extraterrestrische Physik,
  Giessenbachstrasse, Postfach 1312, D-85748 Garching, Germany}

\altaffiltext{8}{NSF Astronomy and Astrophysics Postdoctoral Fellow}

\altaffiltext{9}{Carnegie Fellow, Carnegie Observatories, 813 Santa
  Barbara Street, Pasadena, CA 91101}

\altaffiltext{10}{Departamento de Astronom{\'i}a, Universidad de Chile, 
  Casilla 36-D, Santiago, Chile}

\altaffiltext{11}{Goldberg Fellow, National Optical Astronomy Observatory, 
  950 North Cherry Avenue, Tucson, AZ 85719}

\begin{abstract} 
  Using the Gemini Near-InfraRed Spectrograph (GNIRS), we have
  assembled a complete sample of 20 $K$-selected galaxies at
  $2.0<z<2.7$ with high quality near-infrared spectra. As described in
  a previous paper, 9 of these 20 galaxies have strongly suppressed
  star formation and no detected emission lines. The present paper
  concerns the 11 galaxies with detected \ha\ emission, and studies
  the origin of the line emission using the GNIRS spectra and
  follow-up observations with SINFONI on the VLT. Based on their
  \nt/\ha\ ratios, the spatial extent of the line emission and several
  other diagnostics, we infer that four of the eleven emission-line
  galaxies host narrow line active galactic nuclei (AGNs). The AGN
  host galaxies have stellar populations ranging from evolved to
  star-forming. Combining our sample with a UV-selected galaxy sample
  at the same redshift that spans a broader range in stellar mass, we
  find that black-hole accretion is more effective at the high-mass
  end of the galaxy distribution ($\sim2.9\,\times\,10^{11}
  M_{\odot}$) at $z\sim2.3$. Furthermore, by comparing our results
  with SDSS data, we show that the AGN activity in massive galaxies
  has decreased significantly between $z\sim2.3$ and $z\sim0$. AGNs
  with similar normalized accretion rates as those detected in our
  $K$-selected galaxies reside in less massive galaxies
  ($\sim4.0\,\times\,10^{10} M_{\odot}$) at low redshift. This is
  direct evidence for downsizing of AGN host galaxies. Finally, we
  speculate that the typical stellar mass-scale of the actively
  accreting AGN host galaxies, both at low and at high redshift, might
  be similar to the mass-scale at which star-forming galaxies seem to
  transform into red, passive systems.
\end{abstract}

\keywords{galaxies: active --- galaxies: evolution --- galaxies:
  formation --- galaxies: high-redshift}

\section{INTRODUCTION}

The galaxy population today can very broadly be divided into a
population of star-forming disk galaxies and a population of passive
early-type galaxies. Although this has been known for a long
time, the physics behind the dichotomy in galaxy properties are still
poorly understood. In particular, the model of the formation of
massive, red elliptical galaxies has frequently been modified, driven
by observations. The recent popular hierarchical galaxy formation
models produce galaxies of the required mass, but can not match their
quiescent stellar populations, unless a mechanism -- such as feedback
from active galactic nuclei (AGNs) -- is invoked to stop star
formation at early times \citep[e.g.,][]{gr04,cr06,bo06,ka06}.
 
Understanding the role of AGNs and in particular their feedback
processes in the star formation history (SFH) of a galaxy is one of
today's major challenges. The tight relation between the black hole
mass and bulge velocity dispersion \citep{fm00,ge00} may imply that
black hole accretion is directly related to the formation of its host
galaxy. Moreover, stellar populations of the host galaxies seem to
relate to the strength of the AGN \citep[e.g.,][]{ka03b}. However, the
effects of AGN feedback processes are still poorly understood. The
best example of AGN feedback ``at work'' is the brightest cluster
galaxy Perseus A, which is injecting energy in the intracluster medium
\citep[e.g.,][]{fa03,fa06}. There are several other recent attempts to
constrain the truncation mechanism, which use the large statistical
data-sets available at low redshift. For example, \cite{sc06} derive
an empirical relation for a critical black-hole mass (as a function of
velocity dispersion) above which the outflows from these black holes
suppress star formation in their hosts.

Although the low redshift studies benefit from large, high-quality
surveys and detailed information, they do not enable us to witness the
star formation truncation in more massive galaxies, as stellar
populations studies show that these objects formed most of their stars
at high redshift \citep[e.g.,][and references therein]{dm06}. Recent
attempts to directly witness the AGN feedback process at $z>2$ have
been limited to studies in which galaxies are selected for their
strong nuclear activity. For example, \cite{ne06} argue that an AGN
driven wind is the only plausible mechanism to explain the outflow of
the gas seen in a powerful radio galaxy at $z=2.16$. A more
statistical approach is studying the X-ray properties of AGNs with
redshift. These studies found that AGNs show a top-down behavior, such
that the space density of the more luminous ones peaks at higher
redshift \citep{st03,ue03,ha05}. \cite{he04} claim that this behavior
reflects the decline of the characteristic mass-scale of actively
accreting black holes with redshift. The fact that this behavior is
strikingly similar to what is found for the stellar populations of
galaxies, such that the stars in more massive galaxies are formed at
higher redshift \citep[e.g.,][]{co96,ju05}, could be a clue that the
two are strongly related. In order to relate these two behaviors, and
understand the role of AGNs in the SFH of massive galaxies it is
crucial to study massive galaxy samples at earlier epochs.

At a redshift of $z\sim2.5$ massive galaxies ($>10^{11} M_{\odot}$)
range from starbursting to evolved\footnote{In this paper, ``evolved''
  is shorthand for having a low specific star formation rate
  (SFR/$M_{*}$)} systems
\citep[e.g.,][]{fr03,fo04,la05,re06,pa06,kr06b,wu06}. Thus a massive
galaxy sample in this redshift range may be expected to contain all
evolutionary stages of the process that transforms a massive star
forming galaxy into a red, quiescent system. As massive galaxies are
bright at rest-frame optical wavelengths, a representative massive
galaxy sample can be obtained by selecting at near-infrared
wavelengths \citep[e.g.,][]{fr03,da04b,vd06}. Furthermore, for
detailed information about the star formation and nuclear activity in
the massive galaxies, spectroscopic information is required. As the
average massive galaxy at $2.0<z<3.0$ is faint in the rest-frame UV
\citep[$R_{\rm AB}=25.9$,][]{vd06}, it is beyond the limits of optical
spectroscopy. Thus, if we want to obtain spectroscopic data on a
representative massive galaxy sample at $z\sim2.5$, we need to observe
at NIR wavelengths.

In order to understand the formation of massive galaxies, we are
undertaking a NIR spectroscopic survey for massive galaxies at
$z\sim2.5$ with GNIRS on Gemini-South. Ideally we would study a
mass-limited sample with no regard for luminosity or color. However,
many massive galaxies are too faint for NIR spectroscopy on today's
largest telescopes. Therefore, we study a $K$-selected sample, which
is much closer to a mass-limited sample than an $R$-selected
sample. We note, however, that we may miss massive galaxies with
comparatively high M/L ratios. So far, our spectroscopically confirmed
$K$-selected sample consist of 20 galaxies with $2.0<z<2.7$. Nine of
these galaxies show no emission lines and are characterized by strong
Balmer/4000\,\AA\ breaks. These galaxies are discussed in
\cite{kr06b}. Here we discuss GNIRS and follow-up VLT/SINFONI
observations of the emission line galaxies in the sample. Throughout
the paper we assume a $\Lambda$CDM cosmology with $\Omega_{\rm
  m}=0.3$, $\Omega_{\rm \Lambda}=0.7$, and $H_{\rm 0}=70$~km s$^{-1}$
Mpc$^{-1}$. All broadband magnitudes are given in the Vega-based
photometric system.

\section{DATA}

\taba

\subsection{Sample}

The galaxies studied in this paper are drawn from a spectroscopically
confirmed $K$-selected galaxy sample at $2.0<z<2.7$
\citep{kr06b}. This spectroscopic sample was originally selected from
the optical and deep NIR infrared photometry provided by the
Multi-wavelength Survey by Yale-Chile \citep[MUSYC,][]{ga06,qu06}. The
original sample, selected for $2.0<z_{\rm phot}<2.7$ and $K<19.7$,
contains 26 galaxies all observed with the Gemini Near-InfraRed
Spectrograph \citep[GNIRS, ][]{el06}. Twenty galaxies have a
spectroscopic redshift within the targeted redshift range. Our
$K$-selected sample seems representative for all $K<19.7$ galaxies in
this redshift range, as both a Mann-Whitney U-test and a
Kolmogorov-Smirnov test show that the spectroscopic sample has a
similar distribution of rest-frame $U-V$ colors as the large
mass-limited photometric sample ($>10^{11} M_{\odot}$) by \cite{vd06}
when applying the same $K$-magnitude cut.

Emission lines were detected for eleven of the twenty galaxies. The
nine galaxies without emission lines are presented and discussed in
\cite{kr06b}. We observed ten of the eleven emission line galaxies
with SINFONI \citep{ei03,bo04}, to obtain higher resolution spectra,
and two-dimensional (2D) information on the line emission. For one of
the emission line galaxies we took no follow-up SINFONI spectra, as
this galaxy is already discussed in detail by \cite{vd05}.

\tabb

\subsection{GNIRS spectra}

The original spectroscopic galaxy sample was observed with GNIRS in
2004 September (program GS-2004B-Q-38), 2005 May (program
GS-2005A-Q-20), 2006 January (program GS-2005B-C-12) and 2006 February
(program GS-2006A-C-6). We used the instrument in cross-dispersed
mode, in combination with the short wavelength camera, the 32 line
mm$^{-1}$ grating ($R\sim1000$) and the 0\farcs675 by 6\farcs2 slit. In
this configuration we obtained an instantaneous wavelength coverage of
1.0 -- 2.5 $\rm \mu m$. The integration times for the emission line
galaxies are listed in Table 1. The observational techniques and
reduction of the GNIRS spectra are described in detail by
\cite{kr06a}. For each galaxy we extract a one-dimensional (1D)
original and low-resolution binned spectrum.

We use the low-resolution continuum spectra in combination with the
optical photometry to obtain the integrated stellar population
properties of the galaxies. The near-infrared spectra are flux
calibrated using $JHK$ photometry. The spectra and $UBVRIz$ fluxes are
fit by \cite{bc03} stellar population models with exponentially
declining SFHs, following the technique described in
\cite{kr06a,kr06b}. We assumed a \cite{sa55} IMF and adopted the
reddening law by \cite{ca00}. The redshift was set to the emission
line redshift during fitting. We allowed a grid of 24 ages between 1
Myr and 3 Gyr, 40 values for \av\ between 0 and 4 mag, and 31 values
for $\tau$ (the SFR decaying time) between 10 Myr and 10 Gyr.

GNIRS is uniquely capable of this technique as for $z\sim2.5$ galaxies
the instrument covers the whole rest-frame optical wavelength regime
in one shot, from bluewards of the Balmer break up to 7000 \AA. The
stellar population modeling, driven by the optical continuum break,
yields redshifts as well. This is in particular useful for galaxies
without detected emission lines, such as passive systems. The stellar
population properties of the emission line galaxies studied in this
paper are listed in Table~\ref{cont}.

\figa

\subsection{SINFONI Spectra}

We observed the ten emission line galaxies with near-infrared
integral-field spectrograph SINFONI during two runs, on 2005 December
10-13 (076.A-0464) and 2006 March 3-4 (076.A-0718). The weather
conditions during both runs were fairly stable, with a median seeing
of 0\farcs4 in the NIR. We use the H+K grating ($R\sim2000$,
$\lambda=14500-25000$\AA) over the 8\arcsec $\times$ 8\arcsec\ field
of view (FOV). The FOV is sliced into 32 slitlets, and the spatial
sampling corresponding to this configuration is 0\farcs25 $\times$
0\farcs125. We observed the galaxies according to an ABA'B' on-source
dither pattern. The offset between A and B is half the FOV (4\arcsec)
in the directions perpendicular to the slitlets. The offset between A
and A' is only 1\arcsec. This dither pattern enables an accurate
background subtraction, as we observe empty sky on both sides of the
object in each slitlet.

The raw SINFONI spectra were reduced using custom IDL scripts that
perform the following steps. We start by correcting for the detector
response by dividing by the lamp flats. Next, we determine the
positions of the spectra on the detector for each slitlet from the
``distortion frames''. These frames are constructed by moving an
illuminated fiber in the direction perpendicular to the slitlet over
the FOV. For every slitlet spectrum we determine the position with
respect to the other spectra at each wavelength by tracing the
illuminated fibre. The slitlet lengths, which are different for each
slitlet, are derived from the obtained fiber traces in combination
with the sky emission from the raw object frames.
 
Next, we make a combined bad pixel map for each individual frame, that
identifies cosmic rays, hot pixels and outliers. To identify cosmic
rays, we first have to remove the sky emission. An initial sky removal
is performed by subtracting the average of the previous and successive
frame. The remaining sky residuals are removed by subtracting the
median flux at each wavelength for each slitlet spectrum, using the
derived locations of the spectra on the detector, and masking the
object position. Next, we identify cosmic rays on the obtained images
using L.A.Cosmic \citep{vd01}. We add any remaining 4$\sigma$ outliers
to the map as well. This map is combined with a common bad pixel map,
constructed from flat and bias frames. In what follows the combined
bad pixel map will be transformed in the same way as the science
images.

We return to the raw images and again we perform a simple
sky-subtraction, now using the bad pixel map to reject cosmic-rays and
other defects. Next, we cut and straighten all spectra, using the
derived positions. For each slitlet spectrum, we straighten the
skylines and perform the wavelength calibration in one step, so that
the data are resampled only once. Now, we can accurately remove
remaining sky at each wavelength for each slitlet spectrum, by masking
the object spectrum. The previous steps yield a 3D datacube for each
exposure. Finally, we combine the data cubes of the individual
exposures, using the bad pixel cubes and the offsets. The final cube
is divided by a response spectrum, created from the spectra of AV0
stars, matched in airmass, and reduced in a similar way as the science
objects.

\figb

\subsection{Extraction of one-dimensional spectra}\label{ext}

We extract the 1D spectra in two different ways depending on whether
the \ha\ and \nt\ emission lines are blended due to a strong velocity
gradient. For these galaxies (ECDFS-3694 and ECDFS-10525) we remove
the relative velocity shifts along the spatial direction before
constructing a 1D spectrum, in order to obtain more accurate line
measurements and ratios. While information on the total velocity
dispersion is thereby lost, this procedure is justified for the
purpose of extracting the integrated line fluxes for the analysis
presented in this paper.

For both methods we start by binning the final SINFONI cubes in the
spatial direction, to combine the `spaxels' of 0\farcs125 by 0\farcs25
to spatial elements of 0\farcs25 by 0\farcs25. As we are mainly
interested in studying the spectrum of the line-emitting gas rather
than the continuum emission, we optimize the extraction of the 1D
spectra in the wavelength range around \ha\ and \nt. We make an
average reconstructed image of the data cube in the wavelength region
around these lines. The wavelength region within twice the velocity
dispersion of the lines is included, avoiding wavelengths with strong
OH lines or low atmospheric transmission. Note that the velocity
dispersion is still unknown at this stage, so we have to iterate a few
times to obtain the correct extraction region and the velocity
dispersion. For galaxies with a velocity gradient, we take into
account the relative velocity shift as well to determine which
wavelength region is included

Next, we select all adjacent pixels in the reconstructed image with a
flux exceeding 0.20-0.55 times the flux of the spatial element with
the most signal. This threshold value is dependent on the S/N of the
line emission in the reconstructed images. In case no velocity
gradient is present, the 1D spectra of the selected spatial elements
are combined to form the final 1D spectrum. A 1D noise spectrum is
constructed from the spectra of all spatial elements that do not
exceed this threshold. 

For the galaxies with a velocity gradient, we fit the spectra of all
selected spatial elements following the fitting procedure described in
\S~\ref{line_meas}. To avoid the residuals of skylines as being
interpreted as lines, and to reduce the number of degrees of freedom,
we fix the width to the velocity dispersion of the final spectrum. The
allowed redshift is also constrained to a certain range, depending on
the maximum velocity shift of the galaxy. Note that these latter two
requirements need a few iterations. We measure \ha\ and the two \nt\
lines in the spectra of all selected spatial elements, and retain the
elements for which \ha\ and \nt\ $\lambda$ 6583 have a total
signal-to-noise (S/N) $>5$. We visually inspect the spectra of each
element to check if the lines are correctly interpreted. The
rest-frame 1D spectra of these selected spatial elements are averaged
together to form the final 1D spectrum. We constructed a final noise
spectrum by quadratically adding the rest-frame noise spectra.

Finally, we added the 1D spectra to the previously measured GNIRS
spectra in cases where the S/N is low and the SINFONI data show no
evidence for strong velocity gradients.

\tabc
\figc

\subsection{Line measurements}\label{line_meas}

We obtain emission line ratios, velocity dispersions and equivalent
widths of the emission lines by modelling the extracted 1D
spectra. The $H$ and $K$ spectral bands were fit separately. The $H$
band contains the lines \hb\ and \ot\ $\lambda\lambda$4959, 5007. The
$K$ band covers \ha\ and \nt\ $\lambda\lambda$6548, 6583. We fit
Gaussian models to each set of lines simultaneously, assuming a
similar width for all three lines and one redshift. We note that the
different emission lines may not have similar line width as they can
originate from different physical processes. However, as we will use
the emission line measurements to identify the main contributor to the
line emission, we assume that all lines originate from the same
emission line region. Furthermore, the \nt\ and \ha\ lines have
similar widths for the galaxies for which we could measure the lines
separately. 

We adopt the ratios of transition probablity between the two \nt\
lines and the two \ot\ lines of 0.34 and 0.33 respectively. Thus, for
each fit there are four free parameters: redshift, line width, and the
fluxes of the two emission lines. The line widths were fixed to the
best-fit value obtained from the set of emission lines (either \hb\
and \ot\ $\lambda\lambda$4959, 500 or \ha\ and \nt\
$\lambda\lambda$6548, 6583) with the highest S/N. For the continuum we
use the best fit to the GNIRS low-resolution continuum spectra,
corrected for velocity broadening, assuming that the stellar
dispersion is similar to the gas dispersion. Thus the continuum
directly includes the Balmer absorption.

The 1D spectra are fitted by minimizing the absolute residuals from
the fit weighted by the noise spectrum. This fitting method is
preferred over $\chi^2$ fitting, as it minimizes the influence of sky
lines and strong noise peaks. Errors on the flux measurements were
determined by fitting 500 simulated spectra, that were constructed
from the original spectrum and the photon noise. In these simulations
we also varied the continuum according to the probability distribution
that followed from modeling the GNIRS spectra. In cases where we fix
the width, as derived from a brighter emission line in the same
spectrum, we also vary the assumed width in the simulations according
to the corresponding probability distribution.

For all galaxies except 1030-2026, we start by fitting \ha\ and the
\nt\ lines. We use the obtained redshift, line width and its
probability distribution to fit \hb\ and the \ot\ lines. For 1030-2026
\ha\ is strongly blended with the \nt\ lines, even in the higher
resolution SINFONI spectra. As this galaxy has a clear \ot\
$\lambda$5007 detection in both SINFONI and GNIRS spectra, we swap the
order. Thus for this galaxy we first derive the redshift, the line
width, and its probability distribution from modeling \ot\ and
\hb. Using the derived redshift, line width and its probability
distribution, we can now measure \ha\ and the \nt\ lines.

Figure~\ref{spectra} shows the 1D spectra and best-fit model for all
detected emission lines. The final 1D spectra include the SINFONI
data, or a combination of both SINFONI and GNIRS. As the SINFONI
extraction method is optimized for the line emitting gas, and the
GNIRS for the continuum, the final 1D spectra may not be
representative for the whole galaxy. Nevertheless, a comparison
between GNIRS and SINFONI for the galaxies which could be measured
separately, show that the \wha's are in good agreement: ECDFS-10525:
$W_{\rm H\alpha, GNIRS}=116^{+12}_{-11}$\AA\ and $W_{\rm H\alpha,
  SINFONI}=117^{+13}_{-4}$\AA; 1030-807: $W_{\rm H\alpha,
  GNIRS}=56^{+3}_{-13}$\AA\ and $W_{\rm H\alpha,
  SINFONI}=62^{+13}_{-9}$\AA; ECDFS-3694: $W_{\rm H\alpha,
  GNIRS}=125^{+15}_{-9}$\AA\ and $W_{\rm H\alpha,
  SINFONI}=96^{+17}_{-6}$\AA.

The best-fit equivalent widths, line widths and emission line ratios
are listed in Table~\ref{model}. The values for the emission line
ratios are given when both lines have a $>2\sigma$ detection. In case
only one of the two lines has a $>2\sigma$ detection we give a
$2\sigma$ upper or lower limit. As can be seen in Figure~\ref{spectra}
\ha\ is detected for all galaxies, and \nt\ $\lambda$6583 can be
measured for nine of ten galaxies. Thus, we can determine the value
for \nt/\ha\ for these nine galaxies. For 1030-1531, which has no \nt\
detection we give a 2$\sigma$ upper limit. For only two galaxies both
\hb\ and \ot\ $\lambda$5007 are detected at $>2\sigma$. For these
galaxies the ratio \ot/\hb\ can be measured directly. For four objects
we obtained lower or upper limits from the fitting procedure, as one
of the two lines (\hb\ or $\lambda$5007) had a 2$\sigma$
detection. For example, for galaxy 1030-807 the \hb\ line is detected
at 3 sigma and \ot\ $\lambda$5007 has an upper limit. Thus, we can
derive a 2$\sigma$ upper limit on \ot/\hb, from the 97.5\% cut of the
best-fit \ot/\hb\ values of the 500 simulations. For the remaining
four galaxies the modeling results yielded no limit on \ot/\hb, as
neither of the two lines was detected.

For all galaxies we apply a second method to constrain \ot\
$\lambda$5007/\hb, based on the intrinsic ratio of \ha/\hb\
($=2.76$). We determine a 2$\sigma$ lower limit on \hb\ from the
2$\sigma$ lower limit on \ha. Furthermore, we attenuate \hb\ using the
2$\sigma$ upper limit on the best-fit modeled $A_V$ and assuming extra
extinction towards H\,{\sc ii} regions \citep[factor of 0.44 in
$A_V$,][]{ca97,ca00}. We combine the attenuated 2$\sigma$ lower limit
on \hb\ with the modeled 2$\sigma$ upper limit on \ot\ $\lambda$5007
to derive the 2$\sigma$ upper limit on \ot/\hb. The limits are listed
for each galaxy in the last column of Table~\ref{model}.

The \ha\ line of 1030-2026 is not well fitted in
Figure~\ref{spectra}. This is not caused by a wrong redshift
measurement, as we know the redshift of this galaxy very accurately
from the \ot\ $\lambda$5007 line. There seems to be a second peak at a
rest-frame wavelength of 6600 \AA. Although this ``line'' falls on top
of a strong OH line, the emission feature looks real in the 3D
cube. It could well be \ha\ emission from a companion galaxy or a star
forming region located in the outer parts of the galaxy.

\subsection{Line maps}\label{linemaps}

For each galaxy we make reconstructed images of the continuum and line
emission separately. The continuum images includes all observed H+K
emission, excluding wavelengths with low atmospheric transmission or
strong sky emission. The linemaps include both the \ha\ and the \nt\
$\lambda$6583 emission. The S/N is not high enough to make linemaps of
these two components separately. We included the line emission within
two times the velocity dispersion of the emission lines, accounting
for a velocity gradient if present. The continuum emission is removed
using the flux measurements in the wavelength range surrounding the
lines. Finally, we convolve the line map by a boxcar of 3 pixels
(0\farcs375), to smooth out the noise.

Maps of the continuum and line emission are presented in
Figure~\ref{linemap}. The spatial sampling of the reconstructed images
is 0\farcs125 by 0\farcs25, unlike the reconstructed images used to
extract the spectra, which are binned to 0\farcs25 by 0\farcs25. The
area over which the 1D spectra are extracted is about similar to what
is included by the outer contours.

\section{ORIGIN OF THE LINE EMISSION} \label{sec3}

A key goal of this paper is to elucidate the origin of the \ha\
emission detected in slightly over half of the massive galaxies in our
sample. In the local universe, \ha\ emission is usually an indicator
of star formation, and the luminosity of \ha\ directly correlates with
the instantaneous star formation rate \citep{ke98}.  However, other
processes can also ionize Hydrogen, in particular hard radiation from
a central active nucleus. In this section we use different indicators
for star formation and AGN activity to determine the nature of the
population of emission-line galaxies.
 
\subsection{Emission line ratios} \label{io}

Local star-forming galaxies in the SDSS follow well-defined tracks in
diagnostic diagrams featuring various emission line ratios
\citep[e.g.][]{ba81,vo87}. For our sample, the appropriate diagram is
\ot\ $\lambda$5007/\hb\ versus \nt\ $\lambda$6583/\ha\, as these lines
are relatively strong and are covered in both our GNIRS and our
SINFONI spectra. This diagram is shown in Figure 3. Galaxies for which
lines originate from photo-ionization by young stars in H\,{\sc ii}
regions fall on the well-defined sequence. Local galaxies outside this
metallicity driven sequence are dominated by other ionization sources,
typically photo-ionization by a hard spectrum such as produced by an
AGN.
 
\cite{ka03b} empirically separates the AGNs from the H\,{\sc ii}
sequence by the solid line in Figure~\ref{ratios}. The extreme
starburst classification line derived by \cite{ke01} is presented by
the dashed line. The galaxies between these two dividing lines are
classified as composite H\,{\sc ii}-AGN galaxies by
\cite{ke06}. However, both \cite{er06a} and \cite{sh05} suggest that
the H\,{\sc ii} driven sequence is offset to the right at high
redshift. Thus, we have to be careful when applying the classification
scheme as derived for local galaxies to our high-redshift sample, as
the behavior of the ionization ratios at high redshift is not well
understood. The galaxies outside the H\,{\sc ii} sequence are
generally divided into LINERs and Seyfert 2s. The dotted line in
Figure~\ref{ratios} is the dividing line by \cite{ka03b}.

Our eleven emission line galaxies are indicated by the red filled
squares (specific SFR $<0.05$ Gyr$^{-1}$) and blue filled circles
(specific SFR $>0.05$ Gyr$^{-1}$) in Figure~\ref{ratios}. Remarkably,
only four objects fall in the region of galaxies with pure star
formation as defined by \cite{ke06}, while the other seven galaxies
are classified as AGN or composite H\,{\sc ii}-AGN. Three of these
galaxies fall in the Seyfert 2 regime, three galaxies may be LINERs or
Seyfert 2s, and one is classified as a LINER.

While Seyfert 2s are generally accepted as AGNs, the power source of
LINERs is still debated. Although photo-ionization by an AGN is often
the most straightforward explanation, LINER emission has also been
observed in extra-nuclear regions associated with large-scale outflows
and related shocks \citep{ds95,li04}. Shock-producing winds can be
driven by AGNs \citep[e.g.,][]{ce00,ne06}, but also by strong
starbursts \citep[e.g.,][]{li04}.

\figd

\subsection{Other diagnostics of AGN activity}

Given the ambiguity in the interpretation of LINER spectra it is
necessary to examine other indicators of star formation and AGN
activity, to determine for which of the seven galaxies with non
H\,{\sc ii} region-like ratios an AGN is the most likely explanation
(either directly through photo-ionization or through shocks produced
by an AGN-driven outflow).  We first consider which are the most
important additional diagnostics at our disposal: \smallskip

{\em X-ray emission:} As is well known AGNs can be efficiently
identified by their X-ray emission, which is thought to be due to
up-scattered UV photons from the accretion disk. AGN-induced X-ray
emission can be distinguished from that induced by star formation by
the hardness ratio and (particularly) the luminosity. For the galaxies
in SDSS1030 we use XMM data with a depth of 100\,ks (Uchiyama et
al. 2006, in preparation). Three of the ECDFS galaxies (ECDFS-3662,
ECDFS-3694 and CDFS-6036) are in the CDFS proper \citep{gi02}, for
which very deep 1\,Ms Chandra data are available. The other CDFS
galaxies (ECDFS-3896, ECDFS-5754 and ECDFS-10525) are in the
``Extended'' CDFS, for which we use 250\,ks Chandra data
\citep{vi06}. Interestingly, only one out of seven galaxies is
detected. Limits are given in Table~1. These limits indicate that the
AGNs, if present, are either highly obscured or accrete at
sub-Eddington rates.\smallskip

{\em Compactness:} The spatial distribution of the line emission can
provide information on its origin, as narrow line regions of AGNs are
generally compact. However, extended line emission does not rule out
the presence of an AGN for galaxies with active star formation, as
star forming regions contribute to the line emission. Furthermore,
AGNs can produce outflows, which may result in extended line
emission. As can be seen in Fig.~\ref{linemap}, in several galaxies
the line emission is very compact whereas in others it is
extended. \smallskip

{\em H$\alpha$ equivalent width:} As we have estimates for the SFR in
these galaxies from the stellar continuum fitting, we can compare the
observed strength of \ha\ to that predicted from the best-fit
synthetic spectrum. An excess of \ha\ emission could indicate another
ionization source than star formation, such as AGN activity. We
compute the predicted \ha\ line luminosity using the SFR from the
best-fit model to the observed continuum emission (which includes \av\
as a free parameter) and the \cite{ke98} relation. We then divide this
estimate by the continuum luminosity density around the wavelength of
\ha\ as determined from the best-fit synthetic spectrum and corrected
for the best-fit extinction

In Figure~\ref{wha} we compare the measured and expected \wha. The
relation between the two is dependent on the dust geometry, such that
extra extinction towards H\,{\sc ii} regions would move a galaxy below
the 1-to-1 relation (which is indicated by the dotted line). As it is
unlikely that the attanuation in star forming sites is {\em less} than
the continuum attenuation \citep[see][]{ci05}, a galaxy is not
expected to lie above the dotted 1-to-1 relation. However, in case of
another ionization source such as AGN activity, the measured \wha\
would be higher than the expected. We indeed find that several
galaxies fall {\it above} the expected relation, which implies that an
ionization source other than star formation may be contributing to
line emission as well. \smallskip

{\em Star formation activity:} If a galaxy has a high SFR the current
starburst might produce an outflow which results in AGN-like emission
line ratios. Although this outflow is hard to confirm, the {\em
  absence} of a starburst immediately rules out this scenario, and
infers that ionization ratios are indeed due to AGN activity. To
indicate which galaxies have an engine for such an outflow, we divided
the sample in galaxies with high ({\it blue filled circles}) and low
SFRs ({\it red filed squares}) in Figure 3. \smallskip

Next, we assess for each of the seven candidates whether the line
emission is most likely dominated by an AGN or some other process,
using these criteria and others particular to individual objects. We
stress that in none of the cases it is completely clear-cut: even in
the local universe, with the availability of vastly superior data to
ours, it is often impossible to cleanly identify the relative
contributions of AGNs and star formation to the line emission
\citep[e.g.,][]{fi04,ke06}.  \smallskip

{\it 1030-2026}: This galaxy has Seyfert 2 emission line ratios,
compact line emission, and a low SFR (as implied by the continuum
modeling). Thus, a starburst driven wind is very unlikely for this
galaxy. Furthermore, Figure~\ref{wha} shows that ongoing star
formation, as derived from the stellar continuum, cannot account for
the observed H$\alpha$ emission. Finally, the velocity dispersion of
$\sim$450 km\,s$^{-1}$ of \ot\ $\lambda$5007 may be indicative of an
AGN. We classify this galaxy as an AGN.\smallskip

{\it 1030-2329}: The emission line ratios of this galaxy are
indicative of an AGN, or a composite H\,{\sc ii}-AGN galaxy. The
excess of H$\alpha$ emission in Figure~\ref{wha} supports the presence
of another ionization source. Furthermore, for this galaxy a starburst
driven wind is very unlikely, due to the combination of a low
un-obscured SFR (as implied by the continuum) and compact line
emission.  We include this galaxy in our AGN selection.\smallskip

{\it 1030-2728}: This galaxy has similar supporting diagnostics as
{\it 1030-2329}, but the line emission is not compact. However, the
spectrum shows hints that the \nt\ emission peaks in the central
emission blob in Figure~\ref{linemap}, while the lower-right emission
peak is dominated by \ha\ emission. Unfortunately, the S/N of the
individual line maps is not high enough to draw firm conclusions. As
all other diagnostics indicate an AGN as the dominating ionization
source, we add this galaxy to our AGN selection.\smallskip

{\it ECDFS-3662}: This galaxy probably falls in the composite H\,{\sc
  ii}-AGN region of Figure~\ref{ratios}. Unfortunately, we cannot
discriminate between shock-ionization by a starburst driven outflow
and an AGN, as this galaxy has a high SFR and extended line
emission. Other diagnostics do not provide evidence for an AGN: the
object has no X-ray counterpart (in the 1\,Ms data), and is not
identified as an AGN (nor ULIRG) by \cite{al06} using the infrared
continuum shape. Furthermore, the observed H$\alpha$ emission is
consistent with star formation. Although we cannot rule out the
presence of an AGN, we will not include this galaxy in our AGN
sample.\smallskip

{\it ECDFS-3694}: This galaxy is classified as a composite H\,{\sc
ii}-AGN galaxy in Figure~\ref{ratios}. It has a high SFR, extended
line emission, and shows a large velocity gradient of $\sim $1400
km\,s$^{-1}$. This galaxy falls in the field examined by \cite{al06},
and is not classified as a ULIRG or AGN according to its mid-infrared
SED shape. It is undetected in the 1\,Ms X-ray imaging. A starburst
driven outflow seems the most plausible scenario for this galaxy,
although an AGN cannot be ruled out. Thus, this galaxy will not be
part of our AGN candidate sample.\smallskip

{\it ECDFS-3896}: The emission line ratios for this galaxy are
indicative of an AGN or a composite H\,{\sc ii}-AGN galaxy. This
galaxy has a high SFR derived from the stellar continuum, which is
consistent with the observed \ha\ emission. Nevertheless, a starburst
driven wind is somewhat unlikely, as the line emission shows a very
compact structure and the galaxy shows no velocity gradient.
Furthermore, as will be discussed in the next section, the UV-emission
which boosts the SFR in the model fits, may well be due to continuum
emission by the AGN. Optical spectroscopy is needed to clarify this
situation. As an AGN seems the most plausible cause for the emission
line ratios, we include this galaxy in our AGN sample.\smallskip

{\it CDFS-6036}: This galaxy seems the most convincing Seyfert 2 in
Figure~\ref{ratios}. Nevertheless, \cite{vd05}, who studied this
object in detail, suggest that the ratios are caused by shock
ionization due to a starburst-driven wind. The main evidence is the
extension of the high \nt/\ha\ ratios to the outer parts of the galaxy
and the presence of a strong starburst to drive the
outflow. Furthermore, the galaxy shows a strong velocity field, and
both the \ha\ line emission and the X-ray detection in the soft band
(see Table~1) are consistent with the SFR as derived from the stellar
continuum (see also Figure~\ref{wha}). The rest-frame UV spectrum of
this galaxy does not show any AGN features either \citep{da04a}. The
galaxy has a power-law SED ($\alpha=-1.2$), indicative of an ULIRG or
AGN \citep{al06}. As the available diagnostics are compatible with a
starburst- driven wind, we conservatively do not include this galaxy
in the AGN candidate sample. We note that including this object in the
AGN sample would not change our conclusions.

\subsection{Summary and AGN Fraction}\label{sum}

We find that of the eleven emission line galaxies only four have
emission line ratios consistent with the SDSS star forming sequence
\citep{ka03b}. The immediate implication is that other processes than
``normal'' star formation in H\,{\sc ii} regions play a major role in
massive galaxies at $z\sim2.5$. Determining the nature of these
processes is difficult, as many different physical mechanisms can
produce very similar line diagnostics. Using a variety of indicators
we argue that for four of the galaxies the most likely source of the
\ha\ emission is an AGN. These galaxies are indicated by large solid
circles in Figures 3 and 4, and are labeled ``AGN'' in Figure 2 as
well. They are all narrow line AGNs, as the velocity dispersions of
the emission lines are less than 2000 km\,s$^{-1}$ (Table~3).

According to this classification, the AGN fraction among our total
sample of $K$-selected galaxies at $2.0<z<2.7$ is 20\% (4/20). Due to
the various caveats in the AGN classification this fraction may be
underestimated. First, AGNs are easier to identify in quiescent
systems, than in actively star-forming galaxies, due to the strong
contamination of line emission by H\,{\sc ii} regions. This not only
complicates the emission line diagnostics, but also weakens the
argument of the excess of \ha\ emission compared to the stellar
continua, as this excess is easier to detect in galaxies with low
SFRs. Optical spectroscopy may help to uncover AGNs in star-forming
galaxies, by identifying rest-frame UV high-ionization emission lines
such as C\,{\sc IV} and He\,{\sc II}. But also the MIR continuum shape
and spectral features \citep[e.g.,][]{st06,we06} or very deep SINFONI
spectra to reveal the spatial distribution of \nt/\ha\
\citep{ge06,fo06b,ne06} can be used to identify AGNs in these
galaxies. Furthermore, AGNs may have been missed if they are strongly
obscured (mid-infrared data is available for only a few galaxies), or
are just too faint to be detected. On the other hand, we cannot rule
out shocks produced by a starburst driven wind as the origin of the
line emission in at least one of the four candidates.

\cite{pa06} find an AGN fraction of 25\% among a sample of distant red
galaxies \citep[DRGs,][]{fr03} at $1.5<z<3.0$, using X-ray imaging and
the shape of the mid-infrared continuum.  This result is consistent
with our fraction, given the fact that DRGs make up to 70\% of the
massive galaxy population at $2<z<3$ \citep{vd06}. Our result is also
consistent with the result of \cite{vd04}, who find at least two AGNs
among a sample of six emission line DRGs. Although \cite{ru05} find an
AGN fraction of only 5\% among the 40 DRGs in the FIRES MS\,1054 field
\citep{fo06a}, their criterion of $L_X > 1.2 \times
10^{43}$\,ergs\,s$^{-1}$ would almost certainly not pick up any of the
four AGNs in our study.

Our fraction of AGNs among $K$-selected galaxies is significantly
higher than among UV-selected galaxies. \cite{er06b} identify AGNs in
5 out of 114 UV-selected galaxies, based on the presence of broad
and/or high ionization emission lines in the rest-frame UV spectra,
broad \ha\ lines, or very high \nt/\ha\ ratios \citep[see][]{er06b}.
This fraction is similar to the 5\% found by \cite{re05} among the
full spectroscopic sample of UV-selected galaxies using direct
detections in the 2-Ms {\it Chandra} Deep Field North images, and 3\%
found by \cite{ste02} among a sample of 1000 LBGs at $z\sim3$, using
rest-frame UV spectroscopy. The comparison is complicated as all
studies use different selection criteria. However, this cannot explain
the substantial difference in AGN fraction between the UV and
$K$-selected galaxy samples. We will return to this issue in \S\,4.2.

\fige
\figf

\section{IMPLICATIONS}

In the previous section we identified four AGN candidates among the
eleven emission line galaxies in our $K$-selected sample. In this
section we discuss the nature of the host galaxies of these AGNs, and
what this implies for our understanding of the role of AGNs in the star
formation history of galaxies.

\subsection{Stellar populations of AGN host galaxies}\label{pops}

The low resolution GNIRS spectra and broadband SEDs for all four AGN
candidates are presented in Figure~\ref{SEDS}. The strong optical
breaks for three out of four galaxies imply that the continuum
emission in these galaxies is dominated by stellar light. This is less
clear for ECDFS-3896, as the spectrum of this galaxy shows only a weak
break.  The best-fit stellar population models to the spectrum and the
optical photometry, when assuming that the continuum emission
originates from stars only, are also shown in Figure~\ref{SEDS}. The
corresponding population properties are listed in
Table~\ref{cont}. The spectral continuum of three galaxies (1030-2026,
1030-2329 and 1030-2728) are best fit by evolved stellar population
models with low SFRs, while ECDFS-3896 is actively forming
stars. However, as the continuum emission from the AGN might
contribute significantly, the derived stellar mass and population
properties for this galaxy are quite uncertain.

The median absolute and specific SFRs (SFR per unit mass) of the AGN
host galaxies are 9 $M_{\odot}\,\rm yr^{-1}$ and 0.04 Gyr$^{-1}$
respectively. To examine how the stellar populations of the AGN hosts
compare to those in other galaxies in this redshift range, we divide
the total $K$-selected sample into three classes: the quiescent
galaxies without detected emission lines \citep{kr06b}, the AGNs, and
the remaining emission line galaxies. In Figure~\ref{stacks} we show
the stacked spectra and composite broadband SEDs of these three
classes. The stacked spectra are constructed from the individual GNIRS
spectra, normalized over the continua between a rest-frame wavelength
of 4000 and 6000 \AA. We applied a noise weighted stacking method to
avoid the influence of sky lines and wavelength regions with bad
atmospheric transmission. The broadband SEDs are normalized at a
rest-frame wavelength of 5000 \AA. We also show the average of the
individual best-fit stellar population models to the spectra and
broadband SEDs in the lower panels. Note that for the quiescent
galaxies the redshifts are derived from the continuum shape. As we
have less precise redshift measurements for these galaxies, spectral
features will be smoothed out.

Figure~\ref{stacks} illustrates that the average spectrum and
broadband SED of the AGN hosts is intermediate between the
star-forming and quiescent galaxies, although more similar to the
latter. However, this result may be biased, as AGNs are easier to
identify in galaxies with low SFRs (see \S~\ref{sum}). Thus, we cannot
draw firm conclusions on the stellar populations of the full class of
AGN hosts. Nevertheless, the low SFRs in three of the AGN host
galaxies may suggest that the AGN activity is related to the
suppression of the star formation.

\figg

\subsection{Stellar masses of AGN host galaxies}\label{sec_mass}

In order to asses why the AGN fraction among the UV-selected galaxies
by \cite{er06a} is much lower than among our $K$-selected sample, we
plot the $R-K$ and the specific SFR versus the stellar mass for both
samples in Figure~\ref{masses}, indicating the AGNs in black.
Remarkably, the stellar masses of the AGN host galaxies in both
samples are similar: 2.8$\times$10$^{11}M_{\odot}$ and
3.0$\times$10$^{11}M_{\odot}$ for the UV- and $K$-selected galaxies
respectively (for a Salpeter IMF). As the galaxies in the $K$-selected
sample all have similar stellar masses, an under- or overestimation of
the AGN fraction will not change this result significantly. This
preference for massive galaxies may be the main reason why AGNs are
more common among $K$-selected than among UV-selected galaxies.

The apparent mass dependence of AGN host galaxies is affected by
selection effects. The difference in $R-K$ color and specific SFRs
among the $K$-selected and UV-selected AGN host galaxies is most
likely caused by the different sample selections, as three of our AGN
hosts are too faint in the rest-frame UV to be picked up by the BM/BX
technique \citep{st04}. Three out of seven AGNs in
Figure~\ref{masses}a have $R-K>5.2$. Due to the limitations of both
surveys, we have no information about galaxies with $R-K>5.2$ and
stellar masses below 10$^{11} M_{\odot}$, and it could be that AGNs
reside in these galaxies as well. However, even if we ignore all
galaxies with $R-K>5.2$ a strong mass-dependency is apparent in
Figure~\ref{masses}a.

One can argue that this stellar mass dependency of AGN host galaxies
simply reflects the decrease of black hole mass ($M_{\rm BH}$) when
going to lower mass galaxies, which would make the AGNs more difficult
to detect relative to the contribution from star formation. We cannot
address this bias fully with currently available data, but we can
roughly estimate the effect by normalizing the star formation rate by
the stellar mass (Fig.~\ref{masses}b). Considering a line of constant
specific SFR, and assuming that the black hole mass scales with the
stellar mass of a galaxy \citep[e.g.,][]{de06}, the line emission
originating from accretion onto the black hole should be equally easy
to detect as the line emission from star formation. However, at a
specific SFR between 0.1 Gyr$^{-1}$ and 1.6 Gyr$^{-1}$ we only find
AGNs in very massive galaxies. This implies that the black hole
accretion at high redshift is most effective at higher masses.

This ignores the effect of variations in extinction, which are known
to occur in these samples. In case star formation is more extincted at
the high mass end, AGN emission might be easier to detect in these
galaxies. We also note that the samples are incomplete in the shaded
areas indicated in Figure~\ref{masses}. Furthermore, the stellar
masses and specific SFR derived for ECDFS-3896 and the UV-selected
galaxies might be affected by continuum emission from the AGN.

\subsection{Downsizing of AGN host galaxies}\label{down}

Deep X-ray studies suggest that the AGN population exhibits cosmic
downsizing, as the space density of AGNs with low X-ray luminosities
peaks at lower redshift than that of AGNs with high X-ray luminosities
\citep{st03,ue03,ha05}. Using the SDSS galaxies \cite{he04} conclude
that this behavior is driven by a decrease in the characteristic mass
scale of actively accreting black holes. As the total stellar mass and
black hole mass are found to correlate \citep[e.g.,][]{de06}, the
stellar mass of AGN hosts may be expected to decrease as well, when
going to lower redshift.

To explore the behavior of the stellar mass of AGN hosts with
redshift, we compare the AGN fractions of galaxies with similar
normalized accretion rates at high and low redshift per stellar mass
bin in Figure~\ref{fractions}. The high-redshift galaxies are
indicated by the symbols with error-bars. For the UV-selected galaxies
we only use the mass-range to the right of the dashed line indicated
in Figure~\ref{masses}b, as we cannot ensure the detection of AGNs at
lower stellar masses. The fraction peaks at $M \sim 2.9 \times 10^{11}
M_{\odot}$, but as explained in the previous subsections, this may be
due to incompleteness effects.

In order to compare our sample with low-redshift AGNs with a similar
normalized accretion rate we impose a threshold for $L_{\rm
  [NII]}/M_*$, assuming that the black hole mass scales with the
stellar mass. For the low-redshift galaxies we use the same SDSS
galaxy sample and AGN classification by \cite{ka03b}, as plotted in
Figure~\ref{ratios}. We required a $>3\sigma$ detection for both \ha\
and \nt. In case \ot/\hb\ could not be determined, we selected AGNs by
log(\nt/\ha)$>-0.2$. We set the normalized accretion threshold to
the minimum luminosity of all detected [N\,{\sc ii}] lines in our
sample -- as AGNs with these luminosities would have been detected --
divided by the median stellar mass of the AGN hosts in our sample.

At low redshift the accretion rate is often parametrized by the
luminosity of [O\,{\sc iii}] \citep[e.g.,][]{he05}. As [O\,{\sc iii}]
is not detected for most of our galaxies, we use the luminosity of
[N\,{\sc ii}] instead. For AGNs in the local universe there is a broad
positive correlation between the two (see Figure 3), although this
correlation is different for LINERs and Seyfert 2s. Nevertheless,
[N\,{\sc ii}] is acceptable for our purpose, as we only use the line
luminosity to determine a normalized accretion threshold to compare
similar samples at low and high redshift, which include both LINERs
and Seyfert 2s.

Figure~\ref{fractions} shows that SDSS AGN hosts with a similar
normalized accretion rate as our high-redshift AGNs reside in lower
mass galaxies ($\sim4 \times 10^{10} M_{\odot}$). This finding
provides direct observational support that cosmic downsizing of AGNs
is related to the decrease of the characteristic stellar mass of their
host galaxies \citep[see also][]{he04}. This conclusions is fairly
robust, and is not strongly dependent on the threshold value $L_{\rm
  [NII]}/M_*$, or whether we would use a threshold based on the
luminosity of [O\,{\sc iii}] instead: a factor 2 or 3 difference
barely changes the stellar mass at which the AGN fraction peaks at low
redshift. The main concern for using the [O\,{\sc iii}] or [N\,{\sc
  ii}] emission lines as indicators for the accretion rate is the
contamination by star formation. However, in case [N\,{\sc ii}] at
$z\sim2.3$ is more contaminated by star formation than at $z\sim0$,
our normalized ``threshold accretion rate'' would be overestimated,
and we would select less galaxies at low redshift, and find a slightly
lower typical stellar mass.

Figure~\ref{fractions} also shows the total AGN fraction (i.e.,
including fainter AGNs) among the SDSS galaxies versus stellar
mass. This distribution, which is dominated by faint AGNs, peaks at
higher stellar masses. This implies that at similar stellar masses of
the host galaxies, AGNs in massive galaxies at low redshift are less
powerful than those at high redshift. These low-redshift AGNs may be
the low luminosity descendants of the AGNs observed in our
$K$-selected galaxies.

Both these conclusions would be qualitatively similar if our AGN
fraction would be slightly over- or underestimated, as the galaxies in
our sample span only a small range in stellar mass.

\figh
\figi

\subsection{AGNs and the suppression of star formation}
  
The decrease of the stellar mass of actively star-forming galaxies
with redshift \citep{co96,ju05} may imply that the stellar mass at
which the star formation is suppressed also decreases over cosmic
time. \cite{bu06} introduce a critical ``quenching mass'', which
declines significantly between a redshift of 1.4 and 0.5, and drops
even further at low redshift, as the galaxy bi-modality in the SDSS
breaks down at a stellar mass of $3\,\times\,10^{10} M_{\odot}$
\citep{ka03a}. Also, theoretical studies derive critical quenching
masses, above which cooling and star formation are shut down abruptly
\citep[e.g.,][]{ca06}.

Due to incompleteness of our sample, especially at lower stellar
masses, we cannot derive a critical quenching mass for the observed
redshift range. Nevertheless, Figure~\ref{colmag} illustrates that in
contrast with low redshift, the high-mass end at $z\sim2.3$ consists
of both star-forming galaxies and red, quiescent systems. Moreover,
the red, passive systems may dominate the galaxy population beyond
$2.5\,\times\,10^{11} M_{\odot}$ (see Figure~\ref{colmag}). This may
imply that the transition from star-forming to quiescent galaxies at
$z\sim2.3$ occurs at the observed mass range.

Remarkably, at both low and high redshift, the fraction of actively
accreting AGNs may peak at a similar stellar mass range at which the
suppression of the star formation seems to occur \citep[see
also][]{he04}. This may imply that the critical quenching mass tracks
the evolution of the typical stellar or black hole mass at which
active accretion takes place. This is illustrated in
Figure~\ref{colmag}, which shows the rest-frame $u-g$ color and
$D_n(4000)$ versus the stellar mass for both our $K$-selected sample
and the 28 000 low-redshift SDSS galaxies with estimated co-moving
distances in the range $10<d<150\, \rm Mpc~h^{-1}$ by \cite{bl05}
(this is a sub-sample of the SDSS galaxies presented in
Figures~\ref{ratios} and \ref{fractions}).  $D_n(4000)$ measures the
ratio of the average flux density $F_{\nu}$ in the bands 4000-4100
\AA\ and 3850-3950 \AA\ around the break \citep{ba99}. The actively
accreting low-redshift AGNs are identified by the orange contours.

The AGN hosts at low and high redshift seem to have similar
$D_n(4000)$ and rest-frame $u-g$ colors. However, as explained in
Section~\ref{sum}, AGNs in star forming galaxies may have been missed,
and this would lower the typical $D_n(4000)$ and rest-frame $u-g$
colors of AGN host galaxies. We note however, that the AGN
classification in the SDSS probably suffers from a similar
effect. \cite{sa04} and \cite{na06} find a similar location for AGNs
in color space -- in the ``green valley'' between the red sequence and
blue cloud -- for optically and X-ray selected AGNs at $0.5<z<1.1$ and
$0.6<z<1.4$ respectively. Although the AGN host galaxies at both low
and high redshift and the evolved galaxies at high redshift have
colors comparable to the red-sequence galaxies at low redshift, their
4000 \AA\ breaks are much smaller (Fig.~\ref{colmag}). The red colors
of these galaxies mainly reflect their strong Balmer breaks (instead
of the 4000\AA\ break), indicative of a post-starburst SED.

In summary, our results may suggest that the suppression of star
formation is related with an AGN phase. Different models based on the
CDM theory use various types of AGN feedback to stop the star
formation in massive galaxies. \cite{ho06} propose a quasar mode which
is very effective at high redshift, while \cite{cr06} introduce a
radio mode which is more efficient at low redshift and can also
account for the maintenance of ``dead'' galaxies. Our result might be
more consistent with the mechanism proposed by \cite{ho06}, as the
radio mode by \cite{cr06} is not very effective at $z>2$. In the
\cite{ho06} picture, the observed emission lines may be the fading
activity of a preceding quasar phase.

\section{SUMMARY AND CONCLUSIONS}
 
In order to understand the formation of massive galaxies at
high-redshift, we are conducting a NIR spectroscopic survey of massive
galaxies at $z\sim2.5$. Our $K$-selected galaxy sample contains 20
spectroscopically confirmed galaxies at $2.0<z<2.7$. For nine of these
galaxies no emission lines are detected, indicating that the star
formation in these systems is already strongly suppressed
\citep{kr06b}. In this paper we focus on the emission line galaxies in
the sample, and attempt to constrain the main origin of the ionized
emission. Remarkably, we find that at least four of the 20 galaxies in
our sample may host an AGN. These four AGN candidates are identified
by a combination of indicators, such as the emission line ratios,
compactness of the line emission, and the star formation rate of
galaxies as derived from the stellar continua. Furthermore, there are
three additional galaxies with \nt/\ha\ ratios that may be indicative
for AGNs. However, this hypothesis is not supported by the other
diagnostics (compactness of line emission, \wha, MIR spectral shape),
and other ionization mechanisms, like starburst driven winds also seem
plausible for these galaxies.

This work has several direct implications for high redshift
studies. First, as for a significant part of the massive galaxies at
$2.0<z<2.7$ the line emission is not dominated by pure photoionization
processes from H\,{\sc ii} regions, the \ha\ luminosity is not a good
diagnostic for the current star formation rate. Second, none of the
AGN candidates has a counterpart in the 100 ks (for SDSS1030) or 250
ms (for ECDFS) X-ray images. This implies that optical emission line
ratios provide a complementary approach to identify (relatively low
luminosity) AGNs \citep[see also][for low-redshift AGNs]{he05}.

The stellar populations of the four AGN hosts range from evolved to
star forming, with a median absolute and specific SFR of
$9\,M_{\odot}\,\rm\,yr^{-1}$ and 0.04 Gyr$^{-1}$
respectively. However, this result may be biased, as AGNs are easier
to identify in evolved galaxies. 

Combining our $K$-selected sample with a UV-selected sample in the
same redshift range, that spans a much broader range in stellar mass,
we find that black-hole accretion is more effective at the high-mass
end of the galaxy distribution ($\sim2.9\,\times\,10^{11} M_{\odot}$)
at $z\sim2.3$. This result may be partly due to selection effects, as
both samples suffer from incompleteness, and AGNs with the same
normalized accretion rate are easier to identify in more massive
systems.

The AGNs in low redshift SDSS galaxies with the same stellar mass as
our $K$-selected galaxies are less luminous (as implied from their
\nt\ line luminosities) than their high-redshift analogs. These AGNs
may be the low-luminosity descendants of those we detect at
$z>2$. However, AGNs with similar normalized (to $M_{\rm *} \sim
M_{\rm BH}$ ) accretion rates as the $z>2$ AGNs reside in less massive
galaxies ($\sim 4\times10^{10} M_{\odot}$) at low redshift. This is
direct observational evidence for downsizing of AGN host galaxies. It
will be interesting to see when the AGNs in the $z\sim2.3$ massive
galaxies switch off. We note that \cite{ba05} show that AGNs in more
optically luminous, and thus probably more massive host galaxies are
switching off between z=1 and the present time.

In contrast to what we see at low redshift, the massive galaxy
population at $z\sim2.3$ is very diverse. While the red sequence is
starting to build up, half of the massive galaxies are still
vigorously forming stars. The wide spread of properties observed in
massive galaxies at this redshift range, suggests that we sample
different evolutionary stages of the processes that transform a blue
galaxy into a red one. Thus, the transition may take place at the same
mass range at which the AGN fraction may peak. This is similar to what
we see at low redshift: the fraction of AGNs with the same normalized
accretion rate as those observed at high redshift peaks at the same
stellar mass range at which the color bi-modality breaks
down. Furthermore, the rest-frame $u-g$ colors and $D_n(4000)$ of the
AGN host galaxies at low and high redshift span the same range, and
show that actively accreting AGNs mainly reside in post-starburst
galaxies. This may suggest that the suppression of the star formation
is related to an AGN phase, and that the typical stellar mass scale at
which this suppression occurs decreases with redshift. Our results are
qualitatively consistent with the AGN feedback mechanism as proposed
by several theoretical studies \citep[e.g.,][]{ho06}. However, the
detailed mechanism is not well understood, and perhaps our
observations can help to constrain the theory.

The fact that three of the AGN host galaxies have a very low SFR may
suggest that feedback processes are important. However, we do not have
direct evidence that the suppression of the star formation is caused
by the AGNs, as the co-evolution of star formation and AGN activity
can simply be explained by the fact that both the AGN and
star-formation are fueled by a large gas supply. In this picture, the
galaxy simply runs out of gas, leading to a fading of the AGN and the
star formation at approximately the same time. To clarify this
situation it is necessary to determine whether massive star-forming
galaxies in the studied redshift range host AGNs as well, and how
their accretion rates relate to the accretion rates of the AGNs in the
post-starburst galaxies. 

We also note that the AGN classification is somewhat uncertain for
each individual object, and we may have under- or over-estimated the
AGN fraction. Our results are qualitatively similar if one or two
galaxies were misclassified (in either direction). However, our
conclusions obviously do not hold if none of the galaxies in the
sample host an AGN. As the physical origin of LINER emission is not
well understood even at low redshift, we cannot exclude the
possibility that the observed emission line ratios at high redshift
originate from physical processes other than black hole accretion.
Deeper data than presented in this paper for a significantly larger
sample, in combination with optical spectroscopy, deep X-ray data and
MIR photometry should address this concern and shed new light on the
question what processes are responsible for transforming a
star-forming galaxy into a red, passive system.\smallskip\\
 
\acknowledgments We thank the referee for thorough and helpful
comments, and Shanil Virani and Carie Cardamone for their help in
obtaining the X-ray upper limits. This research was supported by
grants from the Netherlands Foundation for Research (NWO), and the
Leids Kerkhoven-Bosscha Fonds. Support from National Science
Foundation grant NSF CAREER AST-0449678 is gratefully acknowledged. DM
is supported by NASA LTSA NNG04GE12G. EG is supported by
no. AST-0201667, an NSF Astronomy and Astrophysics Postdoctoral
Fellowship (AAPF). PL acknowledges support from Fondecyt Grant
no. 1040719.




\begin{thebibliography}{}
\bibitem[Alonso-Herrero et al.(2006)]{al06} Alonso-Herrero, A., et
  al. 2006, \apj, 640, 167
\bibitem[Baldwin et al.(1981)]{ba81} Baldwin, J., Philips, M., 
  \& Terlevich, R. 1981, \pasp, 93, 5
\bibitem[Balogh et al.(1999)]{ba99} Balogh, M.L., Morris, S.L., Yee, H.K.C., 
  Carlberg, R.G., \& Ellingson, E. 1999, \apj, 527, 54
\bibitem[Barger et al.(2005)]{ba05} Barger, A.J., Cowie, L.L., 
  Mushotzky, R.F., Yang, Y., Wang, W.-H., Steffen, A.T., \& 
  Capak, P. 2005, \aj, 129, 578
\bibitem[Bonnet et al.(2004)]{bo04} Bonnet, H. et al. 2004, The ESO 
  Messenger 117, 17
\bibitem[Bower et al.(2006)]{bo06} Bower, R.G., et al. 2006, \mnras,
  370, 645
\bibitem[Blanton et al.(2005)]{bl05} Blanton, M.R., et al. 2005, \aj, 129,
  2562
\bibitem[Bruzual \& Charlot(2003)]{bc03} Bruzual, G. \& Charlot, S. 2003, 
  \mnras, 344, 1000
\bibitem[Bundy et al.(2006)]{bu06} Bundy, K, et al. 2006, \apj, 651, 120
\bibitem[Calzetti(1997)]{ca97} Calzetti, D. 1997, \aj, 113, 162
\bibitem[Calzetti et al.(2000)]{ca00} Calzetti, D., Armus, L., Bohlin, R.C., 
  Kinney, A.L., Koornheef, J., \& Storchi-Bergmann, T. 2000, \apj, 533, 682 
\bibitem[Cattaneo et al.(2006)]{ca06} Cattaneo, A., Dekel, A.,
  Devriendt, J., Guiderdoni, B., \& Blaizot, J. 2006, \mnras, 370,
  1651
\bibitem[Cecil et al.(2000)]{ce00} Cecil, G., et al. 2000, \apj, 536, 675
\bibitem[Cid Fernandes et al.(2005)]{ci05} Cid Fernandes, R., Mateus,
  A., Sodr{\'e}, L., Stasi{\'n}ska, G., \& Gomes, J.M., 2005, \mnras,
  358, 363
\bibitem[Cowie et al.(1996)]{co96} Cowie, L.L., Songaila, A., Hu, E., \&
  Cohen, J.G. 1996, \aj, 112, 839
\bibitem[Croton et al.(2006)]{cr06} Croton, D.J., et al. 2006, \mnras, 365, 11
\bibitem[Daddi et al.(2004a)]{da04a} Daddi, E., et al. 2005, \apj, 600, L127
\bibitem[Daddi et al.(2004b)]{da04b} Daddi, E., et al. 2004, \apj, 617,
  746
\bibitem[Desroches et al.(2006)]{de06} Desroches, L.-B., Quataert, E.,
  Ma, C.-P., \& West, A.A. 2006, \mnras, submitted (astro-ph/0608474)
\bibitem[Dopita \& Sutherland(1995)]{ds95} Dopita, M.A., \& Sutherland, R.S.
  1995, \apj, 455, 468
\bibitem[Eisenhauer et al.(2003)]{ei03} Eisenhauer, F. et al. 2003, 
  SPIE 4841, 1548
\bibitem[Elias et al.(2006)]{el06} Elias, J.H., et al. 2006, SPIE 6269, 139
\bibitem[Erb et al.(2006a)]{er06a}  Erb, D.K., Shapley, A.E., Pettini, M.,
  Steidel, C.C., Reddy, N.A., \& Adelberger, K.L. 2006, \apj, 644, 813
\bibitem[Erb et al.(2006b)]{er06b}  Erb, D.K., Steidel, C.C., Shapley, A.E., 
  Pettini, M., Reddy, N.A., \& Adelberger, K.L. 2006, \apj, 646, 107
\bibitem[Erb et al.(2006c)]{er06c}  Erb, D.K., Steidel, C.C., Shapley, A.E., 
  Pettini, M., Reddy, N.A., \& Adelberger, K.L. 2006, \apj\, 647, 128
\bibitem[Fabian et al.(2003)]{fa03} Fabian, A.C., et al. 2003, \mnras,
  344, L43
\bibitem[Fabian et al.(2006)]{fa06} Fabian, A.C., Sanders, J.S.,
  Taylor, G.B., Allen, S.W., Crawford, C.S., Johnstone, R.M., \&
  Iwasawa, K. 2006, \mnras, 366, 41
\bibitem[Ferrarese \& Merritt(2000)]{fm00} Ferrarese, L., \& Merritt, D.
  2000, \apj, 539, L9
\bibitem[Filho et al.(2004)]{fi04} Filho, M.E., Fraternali, F., Markoff, S.,
  Nagar, N.M., Barthel, P.D., Ho, L.C., \& Yuan, F. 2004, \aap, 418, 429
\bibitem[F\"orster Schreiber et al.(2004)]{fo04} F\"orster Schreiber,
  N.M.,  et al. 2004, \apj, 616, 40
\bibitem[F{\"o}rster Schreiber et al.(2006a)]{fo06a} F{\"o}rster Schreiber,
  N.M., et al. 2006a, \aj, 131, 1891
\bibitem[F{\"o}rster Schreiber et al.(2006b)]{fo06b} F{\"o}rster Schreiber,
  N.M., et al. 2006b, \apj, 645, 1062
\bibitem[Franx et al.(2003)]{fr03} Franx, M. et al. 2003, \apj, 587, L79
\bibitem[Gawiser et al.(2006)]{ga06} Gawiser, E. et al. 2006, \apjs, 162, 1
\bibitem[Gebhardt et al.(2000)]{ge00} Gebhardt, K., et al. 2000, \apj,
  539, L13
\bibitem[Genzel et al.(2006)]{ge06} Genzel, R., et al. 2006, Nature,
  442, 786
\bibitem[Giacconi et al.(2002)]{gi02} Giacconi, R., et al. 2002, \apjs,
  139, 369
\bibitem[Granato et al.(2004)]{gr04} Granato, G.L., De Zotti, G.,
  Silva, L., Bressan, A., \& Danese, L. 2004, \apj, 600, 580
\bibitem[Hasinger et al.(2005)]{ha05} Hasinger, G., Miyaji, T., \&
  Schmidt, M. 2005, \aap, 441, 417
\bibitem[Heckman et al.(2004)]{he04} Heckman, T.M., Kauffmann, G.,
  Brinchmann, J., Charlot, S., Tremonti, C., \& White, S.D.M 2004, \apj,
  613, 109
\bibitem[Heckman et al.(2005)]{he05} Heckman, T.M., Ptak, A.,
  Hornschemeier, A., \& Kauffmann, G., 2005, \apj, 619, 35
\bibitem[Hopkins et al.(2006)]{ho06} Hopkins, P.F., Hernquist, L.,
  Cox,T.J., Robertson, B., \& Springel, V. 2006, \apjs, 163, 50
\bibitem[Juneau et al.(2005)]{ju05} Juneau, S., et al. 2005, \apj, 619, L135
\bibitem[Kang et al.(2006)]{ka06} Kang, X., Jing, Y.P., \& Silk, J. 2006,
  \apj\ in press (astro-ph/0601685)
\bibitem[Kauffmann et al.(2003a)]{ka03a} Kauffmann G., et al, 2003, 
  \mnras, 341, 33
\bibitem[Kauffmann et al.(2003b)]{ka03b} Kauffmann G., et al. 2003,
  \mnras, 346, 1055
\bibitem[Kennicutt(1998)]{ke98} Kennicutt, R.C. 1998, ARA\&A, 36, 189
\bibitem[Kewley et al.(2001)]{ke01} Kewley, L.J., Dopita, M., Sutherland, R.,
  Heisler, C., \& Trevena, J. 2001, \apj, 556, 121
\bibitem[Kewley et al.(2006)]{ke06} Kewley, L.J., Groves, B., Kauffmann, G.,
  \& Heckman, T. 2006, \mnras, 372, 961
\bibitem[Kriek et al.(2006a)]{kr06a} Kriek, M., et al. 2006a, 
  \apj, 645, 44
\bibitem[Kriek et al.(2006b)]{kr06b} Kriek, M., et al. 2006b, 
  \apj, 649, L71
\bibitem[Labb\'e et al.(2005)]{la05} Labb\'e, I., et al. 2005, \apj, 624, L81
\bibitem[L\'ipari et al.(2004)]{li04} L\'ipari, S., et al. 2004, \mnras,
  355, 641
\bibitem[Nandra et al.(2006)]{na06} Nandra, K., et al. 2006, \apj, in
  press (astro-ph/0607270)
\bibitem[Nesvadba et al.(2006)]{ne06} Nesvadba, N.P.H., Lehnert, M.D.,
  Eisenhauer, F., Gilbert, A., Tecza, M., \& Abuter, R. 2006 \apj,
  650, 693
\bibitem[Papovich et al.(2006)]{pa06} Papovich, C., et al. 2006, \apj, 
  640, 92
\bibitem[Quadri et al.(2007)]{qu06} Quadri, R., et al. 2007, \apj,
  654, 138
\bibitem[Reddy et al.(2005)]{re05} Reddy, N.A., Erb, D.K., Steidel, C.C.,
  Shapley, A.E., Adelberger, K.L., \& Pettini, M. 2005, \apj, 633, 748
\bibitem[Reddy et al.(2006)]{re06} Reddy, N.A., et al. 2006, 644, 792
\bibitem[Rubin et al.(2005)]{ru05} Rubin, K.H.R., van Dokkum, P.G.,
  Coppi, P., Johnson, O., Fo\"rster Schreiber, N.M., Franx, M.,
  \& van der Werf, P. 2005, 613, L5 
\bibitem[Salpeter(1955)]{sa55} Salpeter, E.E. 1955, \apj, 121, 161
\bibitem[S\'anchez et al.(2004)]{sa04} S\'anchez, S.F. et al. 2004,
  \apj, 614, 586
\bibitem[Schawinski et al.(2006)]{sc06} Schawinski, K., et al. 2006,
  Nature, 442, 888
\bibitem[Shapley et al.(2005)]{sh05} Shapley, A.E., Coil, A.L., 
  Ma, C.-P., \& Bundy, K. 2005, \apj, 635, 1006
\bibitem[Steffen et al.(2003)]{st03} Steffen, A.T., Barger, A.J., 
  Cowie, L.L., Mushotsky, R.F., \& Yang, Y. 2003, \apj, 596, L23
\bibitem[Steidel et al.(2002)]{ste02} Steidel, C.C., Hunt, M.P.,
  Shapley, A.E., Adelberger, K.L., Pettini, M., Dickinson, M., \&
  Giavalisco, M. 2002, \apj, 576, 653
\bibitem[Steidel et al.(2004)]{st04} Steidel, C.C., Shapley, A.E.,
  Pettini, M., Adelberger, K.L., Erb, D.K., Reddy, N.A., Hunt, M.P.
  2004, \apj, 604, 534
\bibitem[Stern et al.(2006)]{st06} Stern, D., et al. 2006, \apj, submitted
  (astro-ph/0608603)
\bibitem[Tremonti et al.(2004)]{tr04} Tremonti, C.A. et al. 2004, \apj, 
  613, 898
\bibitem[Ueda et al.(2003)]{ue03} Ueda, Y., Masayuki, A., Ohta, K.,
  \& Miyaji, T. 2003, \apj, 598, 886
\bibitem[van Dokkum(2001)]{vd01} van Dokkum, P.G. 2001, \pasp, 113, 1420
\bibitem[van Dokkum(2004)]{vd04} van Dokkum, P.G. et al. 2004, \apj,
  611, 703
\bibitem[van Dokkum et al.(2005)]{vd05} van Dokkum, P.G., Kriek, M., 
  Rodgers, B., Franx, M., \& Puxley, P. 2005, \apj, 622, L13
\bibitem[van Dokkum et al.(2006)]{vd06} van Dokkum, P.G., et al. 2006, 
  \apj, 638, L59 
\bibitem[van Dokkum \& van der Marel(2006)]{dm06} van Dokkum, P.G., \& 
    van der Marel, R.P., 2006, \apj, 655, 30
\bibitem[Veilleux \& Osterbrock(1987)]{vo87} Veilleux, S., \& 
  Osterbrock, D. 1987, \apjs, 63, 295
\bibitem[Virani et al.(2006)]{vi06} Virani, S.N., Treister, E., Urry, C.M.,
  \& Gawiser, E. 2006, \aj, 131, 2373
\bibitem[Weedman et al.(2006)]{we06} Weedman, D., et al. 2006, \apj,
  653, 101
\bibitem[Wuyts et al.(2006)]{wu06} Wuyts, S., et al. 2006, \apj, 655, 51
\end{thebibliography}
\end{document}